\documentclass[english,10pt]{article} 

\usepackage[T1]{fontenc}
\usepackage[utf8]{inputenc}
\usepackage[english]{babel}
\usepackage{lmodern}
\usepackage{times}
\usepackage{float}
\usepackage{array}
\usepackage{xspace}
\usepackage{xparse}
\usepackage{amsmath}
\usepackage{amssymb}
\usepackage{xifthen}
\usepackage{marvosym}
\usepackage{hyperref}
\usepackage{listings}
\usepackage[all]{hypcap}
\usepackage{graphicx}
\usepackage{subfigure}
\usepackage{todonotes}
\usepackage{booktabs}
\usepackage{caption}
\usepackage{xcolor}
\usepackage{tcolorbox}
\usepackage{tikz}
\usepackage{pgfplots}
\usepgfplotslibrary{groupplots}
\usetikzlibrary{patterns}
\usepackage[section, 
symbols,nogroupskip,nonumberlist,sort=use]{glossaries-extra}
\usepackage{glossary-mcols}
\usepackage{glossary-longbooktabs}
\glsdisablehyper
\usepackage{arxiv}

\graphicspath{{./figures/}}

\RequirePackage{mleftright} 
\RenewDocumentCommand{\Pr}{ s o o m }{%
  \IfBooleanTF{#1}{%
    \IfNoValueTF{#2}{%
      \specrandop{Pr}[#4]%
    }{%
      \IfNoValueTF{#3}{%
        \specrandop{Pr}\mathopen{#2[}#4\mathclose{#2]}%
      }{%
        \specrandop{Pr}_{#3}\mathopen{#2[}#4\mathclose{#2]}%
      }
    }%
  }{%
    \IfNoValueTF{#2}{%
      \specrandop{Pr}\mleft[#4\mright]%
    }{%
      \specrandop{Pr}_{#2}\mleft[#4\mright]%
    }
  }%
}

\title{New CAP Reduction Mechanisms for IEEE 802.15.4 DSME to Support
  Fluctuating Traffic in IoT Systems}  
\newcommand{\shorttitle}{New CAP Reduction Mechanisms for IEEE 802.15.4 DSME}
\author{Florian Meyer, Ivonne Mantilla-Gonz{\'a}lez, Volker 
Turau\\\\Institute of Telematics\\Hamburg University of Technology}

\newcommand{\ACAP}{\mbox{ACR}\xspace}
\newcommand{\DCAP}{\mbox{DCR}\xspace}
\newcommand{\CAPR}{\mbox{CR}\xspace}
\newcommand{\NCAPR}{\mbox{NCR}\xspace}
\newcommand{\GTSCAP}{CAP-GTS\xspace}
\newcommand{\GTSCFP}{CFP-GTS\xspace}

\newcommand{\MSF}{\textit{MSF}\xspace}
\newcommand{\SF}{\textit{SF}\xspace}
\newcommand{\BI}{\textit{BI}\xspace}
\newcommand{\GTSRQ}{\textit{GTS Request}\xspace}
\newcommand{\GTSRP}{\textit{GTS Response}\xspace}
\newcommand{\GTSNF}{\textit{GTS Notify}\xspace}

\makenoidxglossaries

\newglossaryentry{niveau1}{name=niveau1,description={\nopostdesc}}

\glsxtrnewsymbol[description={\textbf{DSME Attributes and Parameters}}]{header1}{}
\glsxtrnewsymbol[description={}]{space1}{}

\glsxtrnewsymbol[description={aUnitBackoffPeriod (symbols)}]{DUB}{\ensuremath{\mathit{aUB}}}

\glsxtrnewsymbol[description={CSMA/CA backoff 
	exponent}]{BE}{\ensuremath{\mathit{BE}}}
\glsxtrnewsymbol[description={Beacon interval}]{BI}{\ensuremath{\mathit{BI}}}
\glsxtrnewsymbol[description={Beacon 
	order}]{BO}{\ensuremath{\mathit{BO}}}

\glsxtrnewsymbol[description={Contention access period}]{CAP}{\ensuremath{\mathrm{CAP}}}
\glsxtrnewsymbol[description={Contention free period}]{CFP}{\ensuremath{\mathrm{CFP}}}

\glsxtrnewsymbol[description={Guaranteed time slot}]{GTS}{\ensuremath{\mathrm{GTS}}}

\glsxtrnewsymbol[description={Multisuperframe 
	duration }]{DMSF}{\ensuremath{\mathit{MD}}}
\glsxtrnewsymbol[description={Multisuperframe 
	order}]{MO}{\ensuremath{\mathit{MO}}}

\glsxtrnewsymbol[description={Superframe 
	order}]{SO}{\ensuremath{\mathit{SO}}}

\glsxtrnewsymbol[description={}]{space2}{}
\glsxtrnewsymbol[description={\textbf{Proprietary Parameters}}]{header2}{}
\glsxtrnewsymbol[description={}]{space3}{}

\glsxtrnewsymbol[description={CAPs per 
	multisuperframe}]{NCAP}{\ensuremath{\mathit{CM}}}
\glsxtrnewsymbol[description={CAP's time slots per 
	multisuperframe}]{SCAP}{\ensuremath{\mathit{CTM}}}

\glsxtrnewsymbol[description={GTSs per  
	beacon interval}]{SGTS_BI}{\ensuremath{\mathit{GB}}}
\glsxtrnewsymbol[description={GTSs per  
	superframe}]{SGTS_SF}{\ensuremath{\mathit{GS}}}

\glsxtrnewsymbol[description={Multisuperframes 
	per beacon interval}]{NMSF}{\ensuremath{\mathit{MB}}}

\glsxtrnewsymbol[description={Expected number of 
	time slots to send a CAP message}]{TSLOT}{\ensuremath{N_{\mathrm{CAP}}}}

\glsxtrnewsymbol[description={Probability of generating a 
	packet during the  CAP}]{PCAP}{\ensuremath{p_{\mathrm{CAP}}}}

\glsxtrnewsymbol[description={Superframes per 
	beacon interval}]{NSF_BI}{\ensuremath{\mathit{SB}}}
\glsxtrnewsymbol[description={Symbols per 
	CAP}]{SC}{\ensuremath{SbC}}
\glsxtrnewsymbol[description={Symbols per 
	superframe}]{SSF}{\ensuremath{\mathit{SbS}}}
\glsxtrnewsymbol[description={Superframes per 
	multisuperframe}]{NSF}{\ensuremath{\mathit{SM}}}

\glsxtrnewsymbol[description={Time slots per beacon interval}]{NTS_BI}{\ensuremath{\mathit{TB}}}
\glsxtrnewsymbol[description={Expected time to send a CAP message (symbols)}]{TC}{\ensuremath{T_{\mathrm{CAP}}}}
\glsxtrnewsymbol[description={Expected channel access 
	time (symbols)}]{T}{\ensuremath{T_\mathrm{Ch}}}
\glsxtrnewsymbol[description={Expected channel access time for a packet generated during the CAP (symbols)}]{TCAP}{\ensuremath{T_{\mathrm{ChCAP}}}}
\glsxtrnewsymbol[description={Expected channel access time for a packet generated during the CFP (symbols)}]{TCFP}{\ensuremath{T_{\mathrm{ChCFP}}}}
\glsxtrnewsymbol[description={Time slots per 
	multisuperframe}]{NTS}{\ensuremath{\mathit{TM}}}
\glsxtrnewsymbol[description={Time slots per 
	superframe}]{NTS_SF}{\ensuremath{\mathit{TS}}}

\glsxtrnewsymbol[description={Smoothing parameter of OpenDSME scheduler}]{alpha}{\ensuremath{\alpha}}
\glsxtrnewsymbol[description={Packets per second}]{delta}{\ensuremath{\delta}}
\glsxtrnewsymbol[description={Ratio of superframes to CAPs per multisuperframe}]{OMEGA}{\ensuremath{\eta}}
\glsxtrnewsymbol[description={Fraction of CFP's time slots a dataframe}]{tau}{\ensuremath{\tau}}

\glsaddall

\begin{document}
\renewcommand\sectionautorefname{Section}
\renewcommand\subsectionautorefname{Section}
\renewcommand\subsubsectionautorefname{Section}

\maketitle

\begin{abstract}
  In 2015, the IEEE 802.15.4 standard was expanded by the
  Deterministic and Synchronous Multi-Channel Extension (DSME) to
  increase reliability, scalability and energy-efficiency in
  industrial applications. The extension offers a TDMA/FDMA-based
  channel access, where time is divided into two alternating phases, a
  contention access period (CAP) and a contention free period (CFP).
  During the CAP, transmission slots can be allocated offering an
  exclusive access to the shared medium during the CFP. The fraction
  \gls{tau} of CFP's time slots in a dataframe is a critical value,
  because it directly influences agility and throughput. A high
  throughput demands that the CFP is much longer than the CAP, i.e., a
  high value of the fraction \gls{tau}, because application data is
  only sent during the CFP. High agility is given if the expected
  waiting time to send a CAP message is short and that the length of
  the CAPs are sufficiently long to accommodate necessary
  (de)allocations of GTSs, i.e., a low value of the fraction
  \gls{tau}. Once DSME is configured according to the needs of an
  application, the fraction \gls{tau} can only assume one of two
  values and cannot be changed at run-time. In this paper, we propose
  two extensions of DSME that allow to adopt \gls{tau} to the current
  traffic pattern. We show theoretically and through simulations that
  the proposed extensions provide a high degree of responsiveness to
  traffic fluctuations while keeping the throughput high.
\end{abstract}

\glsfindwidesttoplevelname    
\printnoidxglossary[type=symbols, style=long-booktabs, 
title={List of Symbols}]

\section{Introduction}
\label{sec:introduction}
The IEEE 802.15.4 standard is the leading adopted networking
specification for applications of the Internet of Things (IoT). While
the specification has been primarily designed with the goal of
lowering the energy consumption of devices, recent extensions focus on
improving reliability, robustness and latency. To ensure transmission
reliability and the timeliness of wireless communication various MAC
protocols have been devised. The main objective is to prevent
interference caused by concurrently transmitting devices out- and
inside of the network. A common approach against the former type of
interference is frequency hopping, whereas reservation schemes are a
means against the latter type. The underlying idea of reservations is
to exclusively reserve some resources -- time, frequency, etc. -- for
each node so that no interference occurs. Examples are SDMA, FDMA, and
TDMA schemes. While these schemes can provide a high degree of
reliability they also have some downsides. The main issue is when and
how to perform reservations. They can be made statically or
dynamically and they can be performed in a distributed fashion or by
a central entity.

Static schedules have the disadvantage that the topology and the
traffic characteristics must be known up front. If traffic patterns
change over time, then this can lead to a waste of resources when
reservations are not fully utilized or on the other hand to a
shortage of resources implicating an increase of latency or in the
worst case a buffer overflow and hence, lost packets. To support
fluctuating traffic patterns such as bursts of packets in a timely
manner, reservations must be made dynamically and in a distributed
fashion. To allow all nodes to make new reservations or withdraw
existing ones at all times, IEEE 802.15.4 provides a mode where time is
divided into two parts: {\em contention free period} (CFP) and a {\em
  contention access period} (CAP). The handling of reservations is
done during the CAP while the CFP is divided into time slots of equal
length called {\em guaranteed time slots} (GTS). A pair consisting of a
GTS and a valid frequency can be reserved for a particular link in
such a way that links which have been allotted the same pair are
spatially separated such that no interference occurs. Access to the
channel in the CAP is managed by a CSMA/CA protocol.

In order to have a network-wide understanding of the two periods, time
is organized with a fixed frame structure, where frames set out the
times for the CFP and the CAP. Furthermore, a time synchronization
mechanism is provided that guarantees that all nodes have a common
understanding of the frame structure, i.e., each frame starts for each
node at the same time. Such a scheme called {\em Deterministic
  Synchronous Multichannel Extension} (DSME) was set up in the IEEE 802.15.4e extension \cite{ieee2015dsme}. The big advantage of DSME is
that the slot allocation mechanism transparently guarantees conflict
free schedules, i.e., users are freed from this elaborate task.
Various integral configuration parameters such $\gls{SO},\gls{MO},\gls{BO}$ of DSME
allow to define the length of a slot, the internal structure of a
frame, and the frequency of time synchronization. The values of these
parameters have to be chosen according to the needs of the
application, e.g., size of data packets, latency, etc. See
\cite{Meyer:2019} for a discussion.

The objectives of a high degree of agility and a high throughput are
conflicting. A high throughput demands that the CFP is much larger
than the CAP, because application data is only sent during the CFP.
Thus, The fraction \gls{tau} of CFP's time slots in a
dataframe is a critical value and each application demands its
dedicated fraction. High agility is given if the expected waiting time
\gls{TC} to send a CAP message is short and that the length of the
CAPs are sufficiently long to accommodate the necessary
(de)allocations of GTSs.

Currently, the IEEE 802.15.4 DSME has defined two standard operating modes: CAP reduction, i.e., \CAPR and no CAP reduction, i.e., \NCAPR. The fraction $\gls{tau}$ in \NCAPR mode corresponds to $7/16$ and in \CAPR mode increases to $(15-8(2^{\gls{SO}-\gls{MO}}))/16$, i.e., $\gls{tau}=0.4375, 0.6875, 0.8125$. From these values it is clear that these two operating modes offer extremely different fractions. This does not allow an optimal usage of resources
for a wide spectrum of applications. If the fraction \gls{tau} is high, then there is little time to make new reservations or to withdraw existing
reservations. Thus, high fluctuations are not well supported. On the
other hand, in times with constant traffic rates there is no need for a
large CAP and valuable time lies wasted. The problem is intensified by
the fact that the configuration parameters of DSME cannot be easily changed
dynamically.

The goal of this work is to extend DSME such that the fraction \gls{tau} can
be changed with a fine granularity and that changes can be performed
dynamically, i.e., after the deployment of the network. We present two
new mechanisms that allow more flexibility in setting up the fraction \gls{tau} and
therefore provide a better support for dynamically varying traffic.
The first mechanism -- Alternating \CAPR (\ACAP) -- operates alternating between \CAPR to \NCAPR every \BI and therefore the actual number of CFP's time slots per \BI varies depending on the operating mode that \ACAP is working. Then, \ACAP's \gls{tau} is the mean of fraction \gls{tau}-values of \CAPR and \NCAPR. The second mechanism -- Dynamic \CAPR (\DCAP) -- allocates GTS in CAPs locally according to the
GTS demands of individual nodes. The higher the demand for GTSs the
shorter will be the available CAP in superframes. \DCAP allows in
principle to adopt any fraction \gls{tau}-value between the fraction \gls{tau}-values of the
two modes \CAPR and no \NCAPR. While \ACAP is far easier to implement
and remains within the original standard, \DCAP is more flexible with respect to the fraction of CFP's time slots in a
dataframe.

We demonstrate, through a theoretical analysis and a series of
simulations, that the two approaches considerably expand the
flexibility of DSME. The evaluation was made in terms of packet
reception rate, mean queue length and the maximum number of allocated GTSs. We show that data collection applications with different demands with respect to agility and throughput can be satisfied by choosing the right strategy. We believe that this work considerably broadens the range of applications that can be optimally supported by DSME.

\section{Related Work} \label{sec:related_work}  A comparative
performance analysis of IEEE 802.15.4 and DSME
has been carried out in several
works~\cite{choudhury2020performance,jeong2012performance,kurunathan2017worst,lee2012performance}.
The analysis in \cite{choudhury2020performance}
and~\cite{kurunathan2017worst} also covers the \textit{Time-Slotted
	Channel Hopping} (TSCH) MAC layer. The former work shows that DSME
and TSCH outperform IEEE 802.15.4 in scenarios with real-time
requirements. The study of DSME shows that \CAPR improves latency and throughput in applications with strict demands.
At the same time the energy consumption is higher than \NCAPR mode, since nodes operate in high duty cycles. The latter
work remarks the effectiveness of the multichannel feature of DSME in terms of delay. Furthermore, simulation results demonstrate an
increased network throughput of about 7\% with \CAPR compared to
\NCAPR. 

Jeon et al. evaluate in~\cite{jeong2012performance} single and
multihop topologies. In both cases, the reachable throughput in DSME is higher than IEEE 802.15.4. (e.g. in multihop topology is about twelve times when \CAPR is enabled). A scenario under interference from IEEE 802.11b wireless networks is analyzed in~\cite{lee2012performance}. Mainly, given the channel diversity
capability of DSME, the adaptability to varying traffic load conditions, and robustness, it is demonstrated a better performance of DSME over IEEE 802.15.4. 

A simulative evaluation of DSME is made by F. Kauer in~\cite{Kauer:2019}. Results evidence that for an increasing \gls{MO} the network throughput increases when \CAPR is enabled. However, higher values of \gls{MO} represent a severe reduction in total CAPs, which means that the network is unable to handle the amount of managed traffic. 

Regarding the impact of \CAPR in DSME, Vallati et.al
\cite{vallati2017improving} analyze it from the perspective of network
formation. The authors propose an active backoff mechanism and
appropriate selection of configuration parameters, that along with
\CAPR aim to reduce setup time up to a 60\%. In the same direction, a dynamic multisuperframe tuning technique (DynaMO) in conjunction with \CAPR is proposed in~\cite{kurunathan2020dynamo}. DynaMO was evaluated in openDSME showing a latency reduction up to 15-30\% in a large scale networks.

\section{Overview of DSME}
\begin{figure}
    \centering
    \includegraphics[scale=1.15]{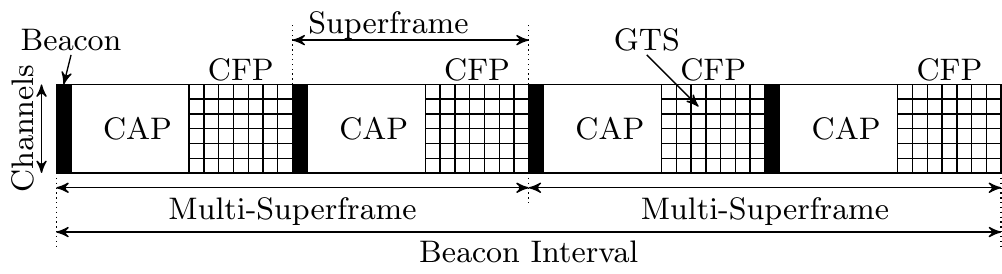}
    \caption{Frame structure in IEEE 802.15.4 DSME 
    ($BO-MO=MO-SO=1$).}
    \label{fig:superframe_structure}
\end{figure}

DSME is a deterministic and synchronous MAC-layer protocol, which
guarantees global synchronization and network parameter dissemination
through beacon messages. The essential parameters are \textit{beacon order} (\gls{BO}), \textit{multisuperframe order} (\gls{MO}) and
\textit{superframe order} (\gls{SO}). Beacons are repeated over time
every \emph{beacon interval} (\BI), with a duration of
$\gls{BI}= 15.36 \times 2^{\gls{BO}}$ seconds. A \BI structures time by
grouping superframes (\SF) into multisuperframes (\MSF) as illustrated
in Fig.~\ref{fig:superframe_structure}. The number of
multisuperframes in a \BI and the number of superframes in a \MSF is
calculated as $\gls{NMSF} =2^{\gls{BO}-\gls{MO}}$ and
$\gls{NSF} = 2^{\gls{MO}-\gls{SO}}$ respectively. A \SF is further
divided in 16 equally sized slots: the first one is for beacon
transmission, the subsequent eight slots form the \textit{contention
  access period} (CAP) and the remaining seven slots the
\textit{contention free period} (CFP).

DSME operates in the 2.4 GHz band and can use 16 channels 
with 5 MHz channel spacing with a transmission rate of 62.5 
ksymbol/s each (i.e. bit rate of 250 kb/s). In the 
CAP, the PAN coordinator selects 
one channel, $C_{\mathrm{CAP}}$, that is used by all network nodes to 
exchange control messages via CSMA/CA. In the 
CFP, nodes communicate through 
\textit{guaranteed time slots} (GTS), which are spread over 
time
and frequency providing exclusive access to the shared 
medium. A schedule of allocated GTSs is repeated every 
\MSF\cite{ieee2015dsme}.

\begin{figure}
	\centering
	\includegraphics[scale=1.15]{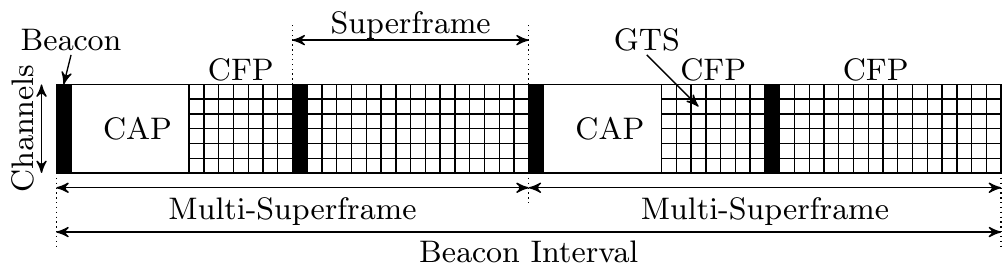}
	\caption{Frame structure in IEEE 802.15.4 DSME with CAP reduction enabled}
	\label{fig:superframe_structure_capOn}
\end{figure}

The available GTS bandwidth per \MSF in \NCAPR mode is
$7\times(2^{\gls{MO}-\gls{SO}}) (\mbox{GTS})$. It can be increased by
enabling the \CAPR mechanism, in which only the CAP of the first \SF
of each \MSF is enabled. Other CAPs are no longer part of the
contention access period but belong to the CFP, as illustrated in
Fig.~\ref{fig:superframe_structure_capOn}. Thus, the total number of
available GTSs per \MSF in \CAPR mode equals
$7 + 15\times(2^{\gls{MO}-\gls{SO}}-1)
(\mbox{GTS})$~\cite{ieee2015dsme}.

\begin{figure}
  \centering
  \includegraphics{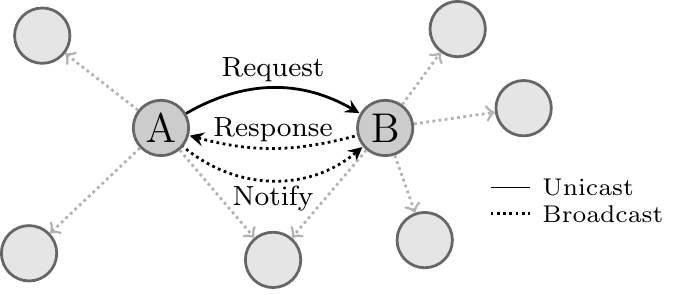}
  \caption{Distributed slot allocation handshake in IEEE 802.15.4 
  DSME.}
  \label{fig:gts_handshake}
\end{figure}

The distributed (de)allocation of a GTS between a neighboring 
pair of nodes is performed in the CAP and follows a 3-way 
handshake as depicted in Fig.~\ref{fig:gts_handshake}. Nodes 
\textit{A} and \textit{B} exchange three messages, a unicast \GTSRQ and two 
broadcasts \GTSRP and \GTSNF. The GTS negotiation guarantees 
a common collision-free selection of the tuple 
(channel, superframe, GTS id). Once neighbors of \textit{A} and \textit{B} 
detect slot inconsistencies by checking their own GTS 
schedules, the allocation is rolled-back and started again.

\subsection{OpenDSME at Glance}

OpenDSME~\cite{Kauer:2019} is an open-source project that implements
features and functionalities of IEEE 802.15.4 DSME in OMNET++. For the
development of new CAP reduction mechanisms the \textit{scheduling
  module} and the management of the DSME data structures
\textit{allocation counter table} (ACT) and \textit{slot allocation
  bitmap} (SAB) are of interest. The first one is a distributed
scheduling algorithm called \textit{Traffic-Aware and Predictive
  Scheduling} (TPS). It is based on an exponentially weighted moving
average filter, which uses a smoothing parameter \gls{alpha} to
estimate the required GTSs with respect to the traffic demand per
link. A hysteresis feature is utilized to enforce a reduction of the
management traffic. Once a link is established between two neighboring
nodes, TPS keeps one slot until the \textit{DSME-GTS expiration} timer
expires. In openDSME a schedule is updated every \MSF. The second
feature of interest is a module that handles GTS status updates when a
GTS negotiation is performed. It ensures the consistency of the ACT
and SAB at every point of the negotiation.


\section{The Proposed CAP Reduction Mechanisms}

The use of \CAPR increases throughput at the cost of reducing the
number of CAPs per \MSF. As noted in ~\cite{Kauer:2019,vallati2017improving}, a severe reduction of
the CAP reduces the responsiveness of the network to varying traffic demands. This effect is intensified for large-scale networks and higher values of \gls{MO} since contention is higher and \gls{TC} is prolonged. Additionally, the adaptability of the network to changing conditions (e.g. external interference) is affected. Therefore, it is necessary to provide enough time in the CAPs for exchanging channel states and based on that to schedule GTSs efficiently~\cite{queiroz2018evaluation}. We propose two CAP reduction mechanisms to address these problems: Alternating CAP-Reduction and Dynamic CAP-Reduction. In the following both mechanisms will be fully described. 

\subsection{Alternating CAP-Reduction (\ACAP)} 
\label{sec:acap}

\ACAP alternates between \NCAPR and \CAPR every beacon interval. The
alternation is not initiated by a central node, e.g., by broadcasting
new configuration parameters into the network. Instead nodes know from
information encoded in the beacons when to switch mode. The advantage
of \ACAP is a higher GTS bandwidth compared to \NCAPR with the minimum
effect on CSMA traffic. That is because switching between \CAPR and
\NCAPR is performed systematically, and nodes do not have to exchange
additional control messages during the CAP to trigger a change of the
operating mode. Moreover, latency does not increase because of the
frame structure realignment after each change of the operating mode
(i.e., from \NCAPR to \CAPR and vice versa).

As part of the initialization, any node except for the root node
(i.e. PAN coordinator), which operates in \ACAP, starts operating in
\CAPR to guarantee that CAP phases will be enabled for CSMA traffic
and will not be affected by self-interference given by TDMA traffic
from nodes already operating in \ACAP.  
Then, after association, nodes
receive a beacon from their parents (e.g. PAN coordinator or
coordinators), in which the cap reduction field is retrieved to
initialize the network's operating mode (i.e. \NCAPR or \CAPR). This
is also done in subsequent {\BI}s. Since nodes know when {\BI}s start,
the implementation of the alternating behavior is straightforward.
Synchronization is guaranteed considering that all beacons transmitted
in the network are allocated within in and repeated every \BI.

\begin{figure}
	\centering
	\includegraphics[scale=1.15]{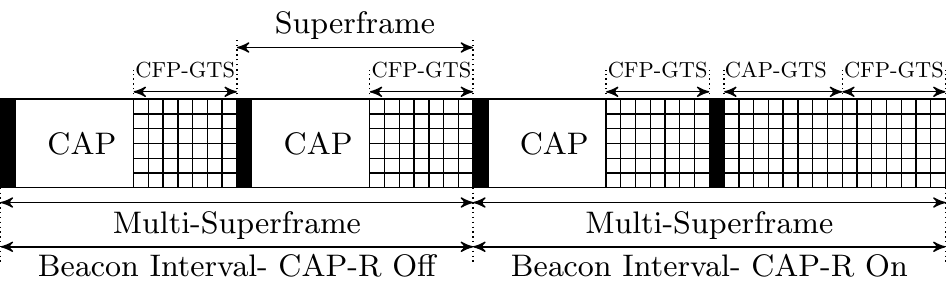}
	\caption{Frame structure in \ACAP ($\gls{BO}=\gls{MO}, 
		\gls{MO}-\gls{SO}=1$).}
	\label{fig:superframe_structure_HCAP}
\end{figure}

Instead of drastically reducing the CAP frequency per \MSF, as the case for \CAPR, \ACAP proposes to enable \CAPR in a way that nodes can autonomously and temporarily increase their throughput or can sleep if no data packets are to be sent. Fig.~\ref{fig:superframe_structure_HCAP} exemplifies the dataframe structure of \ACAP for the case $\gls{MO}= \gls{SO}+1$ and $\gls{BO}= \gls{MO}$. This is, two \textit{SF}s constitute a \MSF and the \BI's length equals that of a \MSF. \ACAP does not increase the potential GTS bandwidth provided by \CAPR. In fact, the fraction \gls{tau} is incremented by the factor $2.14((2^{\gls{MO}-\gls{SO}}-1)/(2^{\gls{MO}-\gls{SO}}+1))$
compared to the fraction \gls{tau} of \NCAPR.

The frame structure of \ACAP is illustrated in
Fig.~\ref{fig:superframe_structure_HCAP} for the case
$\gls{MO}= \gls{SO}+1$ and $\gls{BO}=
\gls{MO}$. 
\ACAP does not increase the potential GTS bandwidth provided by \CAPR.
In fact, \gls{tau} is incremented by the factor
$2.14((2^{\gls{MO}-\gls{SO}}-1)/(2^{\gls{MO}-\gls{SO}}+1))$ compared
to the fraction \gls{tau} of \NCAPR.

\subsubsection{Scheduling of GTSs.}
GTSs in \ACAP are classified according to the type of access period
they belong to: contention access period GTSs (\GTSCAP) or contention
free period GTSs (\GTSCFP). They can be handled in two ways: by defining separate
schedules according to flow priorities or by allocating {\GTSCAP}s as
overprovisioning. The first approach introduces the concept of flows,
that enables higher layers (e.g. routing) to manage different
schedules depending on the priority of packets, i.e., high priority
for flows with a period of $T_{flow} \le \gls{BI}$ and low priority
otherwise. 
The second approach schedules \GTSCAP to deliver queued packets or to
perform packet retransmissions as soon as possible. In this work, the
second alternative is implemented.

\subsubsection{Specifics of \ACAP implementation in openDSME.}
The implementation relies on the management of enhanced beacon
information sent every \BI. Specifically, the CAP reduction field
(i.e. bit 6) contained in the DSME Superframe Specification of the
DSME PAN Descriptor IE. In conjunction with the CAP reduction field,
\ACAP introduces a local Boolean variable in the
\textit{BeaconManager} module to enable the alternating behavior after reception of the first beacon message. Thus, nodes can determine the operating mode for the next \textit{BI}s. Reliability is guaranteed through verification of the operating mode via the cap
reduction field of beacon messages received from the corresponding
parent node. Another implementation aspect is the (de)allocation of GTSs from reduced CAPs, i.e., {\GTSCAP}s. In this work, {\GTSCAP}s are used to deliver queued packets or perform packet retransmissions. Therefore, the scheduler of OpenDSME
makes no difference between \GTSCFP and \GTSCAP to estimate the number of required slots. However, allocation of \GTSCFP has priority over \GTSCAP given their higher frequency. Once all {\GTSCFP}s are allocated, {\GTSCAP}s are negotiated. Deallocation of GTSs follow the opposite order, i.e., a \GTSCAP have a higher priority than a \GTSCFP.

\subsection{CAP-Dynamic Reduction (\DCAP)} 
\label{sec:dcap}
\DCAP starts in \NCAPR mode and is triggered when all GTSs in the CFPs are depleted, i.e., allocated. At this point, \DCAP allocates
additional GTSs during CAPs through the standard DSME 3-way handshake.
For this, two allocating nodes, $v$ and $w$, negotiate a GTS during
the last time slot of a random \SF's CAP, shrinking it from that moment
on. The last slot is chosen because GTS negotiations are triggered at
the start of CAPs in openDSME and thus the first slots of a CAP are
usually busier than the last slots. Choosing a random CAP ensures that
CAPs are reduced evenly and about the same time is available for GTS
negotiations during all portions of a \MSF. Thereby, \DCAP does not
affect the first CAP of a \MSF. The channel $C$, is chosen so that
$C \neq C_{\mathrm{CAP}}$. Therefore, communication during the new GTS does
not interfere with regular CAP traffic. All nodes, except $v$ and $w$,
can use the CAP normally with the restriction that they cannot
communicate with $v$ or $w$ during the allocated GTS. This means,
\DCAP reduces CAPs locally, as illustrated in
Fig.~\ref{fig:example_acap_dcap}.
After the first allocation of \DCAP, node $v$ and node $H_C$ have
reduced their CAP by one slot, while node $w$ can use one less
CAP slot for communication with them but all CAP slots for
communication with other nodes. The rest of the network remains
unaffected. Optionally, $w$ can hold back messages to node
$H_C$ during the allocated GTS and send those during the next CAP to
reduce traffic. Without this optimization, messages to $v$
or node $H_C$ would be lost, since they listen on a different a
frequency. The message would then be repeated in the 
next CAP.

\begin{figure}
    \centering
    \includegraphics[scale=1.15]{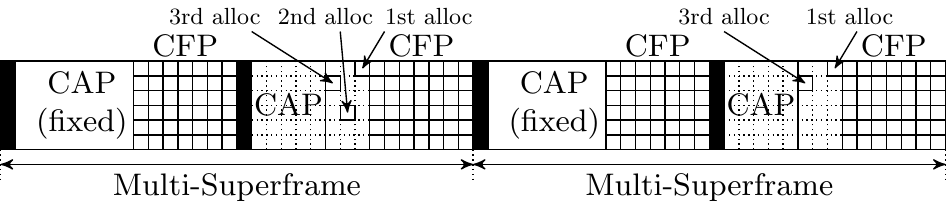}
    \caption{Frame structure using \DCAP ( 
    	$\gls{MO}-\gls{SO}=1$). The allocation
      order for GTSs during the CAP is illustrated.}
    \label{fig:dcap_frame_structure}
\end{figure}

Based on the traffic demand of the network, \DCAP continues allocating
additional GTSs during CAPs until \CAPR\ mode is
reached. That means, the first CAP of every \MSF is not affected to
ensure that GTSs can be deallocated again. This way, up to
$8\times(2^{MO-SO}-1)$ additional GTSs can be allocated. The resulting frame
structure of \DCAP is illustrated in
Fig.~\ref{fig:dcap_frame_structure}, where the first CAP of a \MSF is
immutable, while GTSs can be allocated in the other CAPs, starting with
the last time slot of a CAP. The deallocation of GTSs works exactly the
opposite way until all GTSs in CAPs are deallocated. However, as
illustrated in Fig.~\ref{fig:dcap_frame_structure}, CAPs can become
fragmented during GTS deallocation, e.g., if the first and third GTS
allocation was done by one node and the second GTS allocation was done
by another node. Then, the deallocation of the second GTS results in a
split CAP. A solution for this is the relocation of the third GTS to
a later time slot. This behavior is currently not implemented, but also not mandatory as timers are stopped outside of CAP slots and therefore no timeouts for CAP messages should occur.

\subsubsection{Scheduling of GTSs.}
\DCAP provides opportunities for more sophisticated scheduling
algorithms. For example, a scheduler could decide to allocate
additional GTSs during the CAP if traffic demand is high but
deallocate GTSs again if the PRR of the CAP traffic falls below a
certain threshold, e.g., during traffic fluctuations. Additionally, a
modification of the scheduler has to be
done i.e., the scheduler first allocates GTSs during the CFP until they are
depleted and then starts allocating GTSs during the CAP.

\subsubsection{Specifics of the \DCAP implementation in openDSME.} It must be mentioned that there is a storage overhead for \DCAP  because it requires larger data structures (SAB and ACT) for storing information about the $8\times(2^{\gls{MO}-\gls{SO}}-1)$ additional GTSs. Even if no \CAPR\ is active, nodes must use the SAB and the ACT of the \CAPR mode for storing their GTS information. That is because the size of the structures in DSME cannot be changed dynamically but they are determined during network configuration.
There is no way to estimate if slots will be allocated during the CAP using \DCAP, since the decision is purely based on the current traffic demand of the network, so that the data structures always have to include the potential GTSs during the CAPs. At last, it has to be said that \DCAP is not conform with the IEEE 802.15.4 standard and openDSME requires larger code modifications to allow the allocation of GTSs during the CAP. For example the transceiver cannot be turned on and set to a single frequency for the duration of the whole CAP anymore, but it has to be checked if the next time slot is a GTS and the frequency has to be adapted accordingly. This changes how GTSs are handled. The default behavior at the end of a GTS is to turn off the transceiver. Now, if the GTS was inside a CAP, the frequency has to be switched back to $C_{\mathrm{CAP}}$ and the transceiver has to stay on. This does not increase energy-consumption as the transceiver of nodes must be turned on during the whole CAP. 


\subsection{An example of \ACAP and \DCAP in DSME}
\begin{figure}
    \centering
    \includegraphics{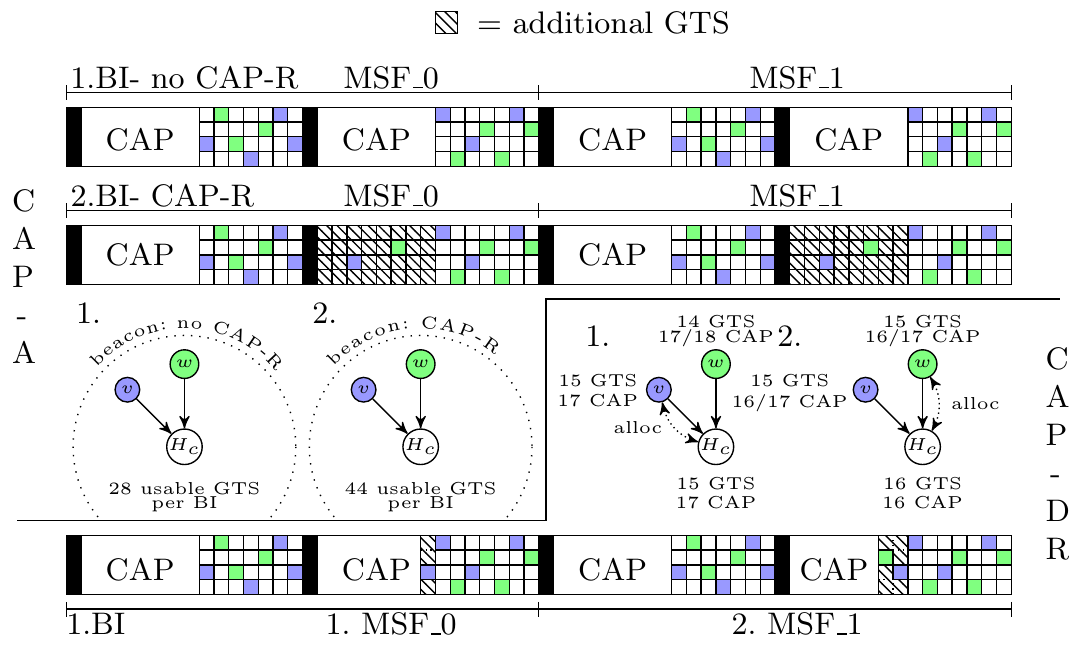}
    \caption{Example of \ACAP and \DCAP making more GTS 
    resources available in a network with 3 nodes ($\gls{BO}-\gls{MO} = 1, 
    \gls{MO}-\gls{SO}=1$).}
    \label{fig:example_acap_dcap}
\end{figure}

Fig.~\ref{fig:example_acap_dcap} shows an example of how the two
proposed CAP reduction mechanisms work. Here two nodes ($v$ and $w$) send messages to the coordinator ($\mbox{H}_{\mbox{c}}$). The
GTS during the CFP are already completely allocated, as illustrated in the schedules for the two mechanisms.

With \ACAP, nodes switch between \CAPR\ and \NCAPR\ every \BI, and
information about the current operating mode is retrieved through
beacons received from $\mbox{H}_{\mbox{c}}$. As it is shown in this
example, during the \BI at time $t$, represented in the first row,
nodes operate in \NCAPR. During that \BI, a total number of 28 GTSs can
be allocated. Then, during the next \BI at time $t+\gls{BI}$, in the
second row, nodes alternate their frame structure to \CAPR\ with a
maximum of 44 usable GTSs. This alternating behavior allows nodes to
allocate extra GTSs in the extended CFPs. \ACAP would even work if the
beacon from $\mbox{H}_{\mbox{c}}$ is not heard by, e.g., $v$
because nodes also keep track of time themselves and can switch
between \CAPR\ and \NCAPR\ autonomously after they heard the first
beacon, immediately after the association to the network is completed.
Additionally, \ACAP allows static schedules because the frame
structure is completely deterministic (i.e. every \BI, the frame
structure switches from \CAPR\ to \NCAPR\ or vice versa).

For \DCAP, additional GTS resources are created through the 
DSME 3-way handshake when all GTSs during CFPs are allocated. 
Here, $v$ allocates an additional GTS with $\mbox{H}_{\mbox{c}}$. For the 
new GTS, the CAPs of $\mbox{H}_{\mbox{c}}$ and $v$ are locally reduced and 
a GTS is allocated on a different channel than $C_{\mathrm{CAP}}$. This 
way, $w$ can continue using all CAP slots for 
communication with nodes other than $v$ and $\mbox{H}_{\mbox{c}}$ (which are not shown here). The benefit is that the number of additionally 
created GTSs is a direct effect of the traffic load and therefore no unused GTS during the CAP occur.

\section{Metrics and Hypotheses} \label{sec:metrics}
There is a number of relevant metrics for the evaluation of the proposed CAP reduction mechanisms. These include, at an abstract level, the adaptability of
DSME with respect to time varying traffic and the fraction
of CFP's time slots in a dataframe, i.e., the fraction \gls{tau}. Adaptability refers the
time for (de)allocating a GTS. That is because the proposed mechanims
try to increase the potential throughput while maintaining
responsiveness and reliability. For the simulative evaluation also the
packet reception ratio (PRR), mean queue length, the maximum number of allocated GTS and dwell time are relevant as they indirectly represent adaptability and throughput of DSME.

It can be expected that overall performance of the described CAP reduction
mechanisms strongly depends on the difference \gls{MO} $-$ \gls{SO},
since the two parameters directly control the number of \SF per \MSF
and thus the frequency of the CAPs per \MSF. An analysis of this
difference can be made considering two cases: Varying \gls{MO} while
fixing \gls{SO} or varying \gls{SO} while fixing \gls{MO}. The former
keeps the slot length constant and for each increment of \gls{MO}, the length of the \MSF as well as the number of CAPs per
\MSF is doubled. The latter preserves the length of the \MSF, and each increment of \gls{SO} doubles the slot length and halves the number of CAPs per \MSF.

In case of \CAPR the CAP frequency per \MSF is always equals 1.
Therefore, for a fixed \gls{SO}, increasing \gls{MO} means enlarging
the \MSF length and thus increasing \gls{TC}. In other words, as the
difference between \gls{MO} and \gls{SO} increases, \CAPR performs
worse as there are not enough CAPs to (de)allocate all GTSs in
time. For a fixed \gls{MO}, decreasing the value of \gls{SO} increases
\gls{NSF} by reducing the length of the \SF and therefore the length
of CAPs. Thus, as the difference between \gls{MO} and \gls{SO}
increases, \CAPR's performance degrades, because the reduction of time
in CAPs diminishes the capability of the network to (de)allocate GTSs
to adapt rapidly to fluctuating traffic.

On the other hand, differences between \gls{MO} and \gls{SO} should
not have an strong effect for \NCAPR. This is, because either fixing
\gls{SO} or \gls{MO}, the number of CAPs per time stays invariant. In
some scenarios (e.g. highly varying traffic), \gls{MO} could be an
influencing parameter because it determines the period in which
the scheduler updates the number of required GTSs to adapt to traffic
changes in the network. Thus, the higher \gls{MO} the lower the
responsiveness of the network to rapid traffic changes over time. In
case of the scheduler of openDSME, it uses hysteresis and a smoothing 
parameter, \gls{alpha}, to manage effectively such traffic
fluctuations in the network. Therefore, the parameters 
\gls{MO} and \gls{SO} should not have a significant 
influence on \NCAPR as stated above. 

In \ACAP, the alternation between \CAPR and \NCAPR every \gls{BI} guarantees that the average amount of time in CAPs is enough to
accomplish the required GTS (de)allocations to meet traffic demands.
Moreover, the scheduling strategy guarantees stability because it
first allocates \GTSCFP and then \GTSCAP, which are less frequent over
time. Therefore, as well as \NCAPR, we do not expect higher changes
in \ACAP's performance given by differences between \gls{MO} and
\gls{SO}. \DCAP should converge to \CAPR or \NCAPR, depending on which
one performs better, and should even exhibit a better performance for
a large difference between \gls{MO} and \gls{SO}. 

\section{Theoretical Evaluation} 
\label{sec:theoretical_evaluation}
The goal of the proposed CAP reduction mechanisms is to increase the throughput while maintaining adaptability to time varying traffic. We consider mainly two metrics:
\begin{enumerate}
	\item   the fraction \gls{tau} and
	\item  the adaptability of DSME expressed in two correlated variables:
	\begin{itemize}
	 	\item the expected time to send a CAP message, \gls{TSLOT}, on time slot level and
	 	\item the expected channel access time, $\gls{T}$, on symbol level, that a node has to wait to perform a GTS negotiation during the next CAP.
	\end{itemize}	 	 
	
\end{enumerate}	

Both metrics strongly depend on the average number of
CAPs per \MSF, $\gls{NCAP}$, and the average number of CAP's time
slots per \MSF, $\gls{SCAP}$. For
example, \CAPR heavily sacrifices adaptability for a higher throughput, which
is especially problematic if not all GTSs are utilized, because they
could have been used for CAP traffic. The proposed mechanisms overcome
this problem by modifying $\gls{NCAP}$ (\ACAP) and $\gls{SCAP}$
(\DCAP).

\subsection{The fraction \gls{tau} of CFP's time slots in a dataframe}
The parameter \gls{tau}, can be calculated based on $\gls{NCAP}$ and
$\gls{SCAP}$ as
\begin{align}
    \tau = \frac{\gls{NTS} - \gls{SCAP} - 
    \gls{NSF}}{\gls{NTS}},  
\end{align}
where $\gls{NTS} = 16 \times \gls{NSF}$ is the total number of time 
slots in a \MSF. Subtraction of \gls{NSF} is required to account 
for \gls{NSF} beacon slots per \MSF. Therefore, the fraction \gls{tau} increases for decreasing values of 
$\gls{SCAP}$, as illustrated in Tbl.~\ref{tbl:time_sending}. 
\NCAPR is independent of \gls{MO}, while \CAPR converges 
towards $93.75\%$ as \gls{MO} 
increases. \ACAP offers a compromise between \NCAPR and 
\CAPR. Similarly, \DCAP dynamically adapts the time for 
sending packets with the lower bound equal to \NCAPR and the 
upper bound equal to \CAPR.

\begin{table}[H]
    \centering
    \begin{tabular}{p{0.12\textwidth}|>{\centering}p{0.2\textwidth}
     >{\centering}p{0.2\textwidth} 
    >{\centering}p{0.2\textwidth} c}
         & $\gls{MO}=4$ & $\gls{MO}=5$ & $\gls{MO}=6$ & 
         $\gls{MO}=7$ \\
        \hline
        \NCAPR & 43.75\% & 43.75\% & 43.75\% & 43.75\% \\
        \hline
        \CAPR & 68.75\% & 81.25\% & 87.5\% & 90.06\% \\
        \hline
        \ACAP & 56.25\% & 62.5\% & 65.63\% & 67.19\% \\
        \hline
        \DCAP & 43.75\% - 68.75\% & 43.75\% - 81.25\% & 43.75\% 
        - 87.5\% & 43.75\% - 90.06\%\\
    \end{tabular}
    \caption{Values of the fraction \gls{tau} for sending packets 
     with \NCAPR, \CAPR, \ACAP, and \DCAP for $\gls{SO}=3$ and $4 \le \gls{MO} \le 7$.}
    \label{tbl:time_sending}
\end{table}


\subsection{ Metrics to evaluate the adaptability of DSME}

The adaptability of DSME is characterized by the expected time to
allocate a new GTS, i.e., the expected time until a GTS-handshake can
be conducted in a CAP. In contrast to \gls{tau}, which indicates the
maximum throughput of a system, adaptability expresses how long it
would take to allocate all required GTSs. Additionally, it provides an insight into how well the system adapts time-varying traffic. Therefore, we define the following two metrics to determine the agility of DSME:

\subsubsection{Expected time to send a CAP 
message on time slot level.}

Since a time slot is a baseline unit in DSME's frame structure, we consider it to estimate adaptability in terms of the expected number of time slots that a node should wait to send a CAP message, i.e. \gls{TSLOT}. This value can be calculated per \MSF as

\begin{align}
    \gls{TSLOT}(\gls{NCAP}, \gls{SCAP}) =
    \frac{\sum_{i=1}^{\frac{\gls{NTS}-\gls{SCAP}}{\gls{NCAP}}}i}{\gls{NTS}}.
\end{align}
The expected number of time slots to send a CAP message in \NCAPR is
$\gls{TSLOT}(\gls{NSF}, 8\times\gls{NSF})$, which is independent of
\gls{MO} with a value of 2.25 time slots. For \CAPR this value is
$\gls{TSLOT}(1, 8)$. Hence, it dependents on the difference between
\gls{MO} and \gls{SO}. As shown in Tbl.~\ref{tbl:channel_access_time},
the expected waiting times to send a CAP message increase
exponentially for an increasing \gls{MO}. For \ACAP, \gls{TSLOT} is
calculated as an average of the expected times for \CAPR and
\NCAPR.
The value amounts to
$0.5 \times (\gls{TSLOT}(\gls{NSF}, 8\times\gls{NSF}) + \gls{TSLOT}(1,
8))$. \DCAP dynamically adjusts the number of additional GTSs between
\NCAPR and \CAPR so that the expected times are bounded by the
expected times of these mechanisms.

\begin{table}
    \centering
    \begin{tabular}{p{0.12\textwidth}|>{\centering}p{0.2\textwidth}
     >{\centering}p{0.2\textwidth} 
    >{\centering}p{0.2\textwidth} c}
        & $\gls{MO}=4$ & $\gls{MO}=5$ & $\gls{MO}=6$
        & $\gls{MO}=7$ \\
        \hline
        \NCAPR & 2.25 & 2.25 & 2.25 & 2.25 \\
        \hline
        \CAPR & 9.38 & 24.94 & 56.72 & 120.61 \\
        \hline
        \ACAP & 5.81 & 14.10 & 29.49 & 61.43 \\
        \hline
        \DCAP & 2.25 - 9.38 & 2.25 - 24.94 & 2.25 - 56.72 &
        2.25 - 120.61 \\
    \end{tabular}
    \caption{Expected time to send a CAP message on time slot level for increasing values of \gls{MO}.}
    \label{tbl:channel_access_time}
\end{table}

\subsubsection{Expected channel access time on 
symbol level.}

The expected channel access time for CAP messages on symbol level, i.e. $\gls{T}$, evaluates the adaptability of DSME as it describes the time until a node is able to send a CAP message and initiate a GTS (de)allocation.
 In the following,  a single channel access without contention is considered where  the corresponding packet is generated according to a uniform  distribution on the interval $[0, \gls{DMSF}]$. Two cases have  to be considered: packet generation during the CFP and packet generation during the CAP. For the calculation, beacon slots are treated as CFP slots in which no data is sent.

\textbf{During the CAP:} The probability, $\gls{PCAP}$, for 
generating a packet during a CAP is given by
\begin{align}
    \gls{PCAP} = \frac{\gls{SCAP} }{\gls{NTS}}.
\end{align}
The system can be discretized in time steps, \gls{DUB}, of 
\textit{aUnitBackoffPeriod}, the base duration for the 
CSMA/CA algorithm. 
If a packet is generated at the end of the CAP, its transmission 
is more likely to be shifted to the next CAP. Therefore, the 
expected CSMA/CA backoff duration, $T_{b}(s)$, for a packet 
generated during the $s$th CSMA/CA slot is given by
\begin{alignat}{2}
    &T_b(s) = \frac{\sum_{i=0}^{i\le 2^{\gls{BE}}-1}  B(s, 
    i)}{  2^{\gls{BE}} - 1 }  \label{eq:tb} \\
    &B(s, i) =  \begin{cases}
                        i\times\gls{DUB}         &\text{ if }\gls{DUB}\times(s+i) < 
                        \gls{SC} \\[1ex]
                        \begin{array}{l}
                            \gls{OMEGA} \times \gls{SSF} - \gls{DUB}\times s\\ + 
                            \gls{DUB}(i+s)\ mod\ \gls{SC}\\
                             + \gls{OMEGA} \times \gls{SSF} \lfloor 
                             \frac{\gls{DUB}(s+i)}{\gls{SC}}\rfloor
                        \end{array}&\text{ otherwise}
                    \end{cases}, \label{eq:B}
\end{alignat}
where $\gls{BE}$ is the backoff exponent of the CSMA/CA 
algorithm, $\gls{SSF} = 16 \times 60 \times 2^{\gls{SO}}$ is the number 
of symbols per \SF,  $\gls{SC} = 
\frac{\gls{SCAP}}{\gls{NCAP}} \times 60 \times 2^{\gls{SO}}$ is the 
number of symbols per CAP, and $\gls{OMEGA} = 
\frac{\gls{NSF}}{\gls{NCAP}}$ is a stretching factor if 
$\gls{NCAP} < \gls{NSF}$ under the assumption that CAPs are 
evenly distributed in the \MSF. In other words: 
Eq.~(\ref{eq:tb}) calculates the expected backoff from a 
given slot for all possible CSMA/CA backoff values. The total 
backoff, $B$, equals to the CSMA/CA backoff ($i\times\gls{DUB}$) if 
there is enough space in the CAP. Otherwise, it includes the 
wait duration of the CFPs. In this case (Eq.~(\ref{eq:B}), case 
otherwise), the first summand is the backoff until the next CAP 
from the time of the packet generation, the second summand is 
the remaining backoff time in a latter CAP, and the third term 
adds the backoff for multiple superframes if $i\times\gls{DUB}$ 
spans over multiple CAP phases. The expected channel access 
time, \gls{TCAP}, for a packet generation during the CAP can 
then be calculated as
\begin{align}
    \gls{TCAP} = \frac{\sum_{s=0}^{s<\gls{SCAP} / 
    (\gls{DUB}\times\gls{NCAP})} 
    T_b(s)}{\gls{SCAP}/(\gls{DUB}\times\gls{NCAP})} .
\end{align}

\textbf{During the CFP:} When a packet is generated during the 
CFP, the expected channel access time is the combination of 
the expected time to send a CAP message, i.e. \gls{TC}, and the expected 
CSMA/CA backoff time from slot 0. The expected time to send a CAP message on symbol level is given by $\gls{TC} = 0.5 \times 
(\gls{TC}_{min} + \gls{TC}_{max})$, where $\gls{TC}_{min} = 1$ is the 
minimum time to send a CAP message and $\gls{TC}_{max}$ is the 
maximum time to send a CAP message and given by
\begin{align}
    \gls{TC}_{max} = \gls{OMEGA} \times \gls{SSF} - \gls{SC}
\end{align}
The expected channel access time, \gls{TCFP}, for a packet 
generated during the CFP is $\gls{TCFP} = \gls{TC}  + 
T_b(0)$. 

The expected channel access time, \gls{T}, is 
then given by
\begin{align}
    \gls{T} = \gls{PCAP} \times \gls{TCAP} + (1-\gls{PCAP}) \times 
    \gls{TCFP}.
\end{align}

\begin{figure}[]
    \centering
    \pgfplotsset{compat=newest}
\colorlet{lightblue}{blue!40}
\pgfplotsset{
	colormap={meins}{color(0cm)=(lightblue);color(0.6cm)=(yellow); color(1cm)=(orange); color(3cm)=(red);}
}

\begin{tikzpicture}
  \begin{axis}[
  		view={128}{26},
  		grid=major,
  		xtick={2,4,6,8,10,12,14,16},
  		ytick={1,2,3,4,5,6,7,8},
  		xticklabel style={xshift=0.1cm, yshift=0.05cm},
  		yticklabels={1, 2, 3, 4, 5, 6, 7, 8},
  		yticklabel style={xshift=-0.05cm, yshift=0.05cm},
  		ztick={0,20,40,60,80,100,120},
  		zticklabel style={yshift=0.2cm}, 
  		xlabel={\gls{NCAP}},
  		ylabel={Avg. \# time slot per CAP}, 
  		ylabel style={rotate=-14.5},
  		zlabel={Expected channel access time [s]},
  		zmin=0,
  	]
    \addplot3 [
    surf, 
    opacity=1.0,
    shader=faceted,
    ] table[x=X, y=Y, z=Z] {plots/access_time_data.csv};
    \addplot3[only marks, mark=*, mark options={fill=green, scale=1.3}, nodes near coords,point meta=explicit symbolic] coordinates{
    	(16,8,2.4166) [\NCAPR]
    	(1,8, 120.61979) [\CAPR]
    	(8.5,8,8.6293) [\ACAP]
      	};
  \end{axis}
\end{tikzpicture}
    \caption{Expected channel access time in time slots for 
    different values of \gls{NCAP} and average number of 
    time slots per CAP for $\gls{SO}=3$ and 
    $\gls{MO}=7$. The performance of the static CAP reduction 
    mechanisms is marked by the green dots.}
    \label{fig:channel_access_time}
\end{figure}
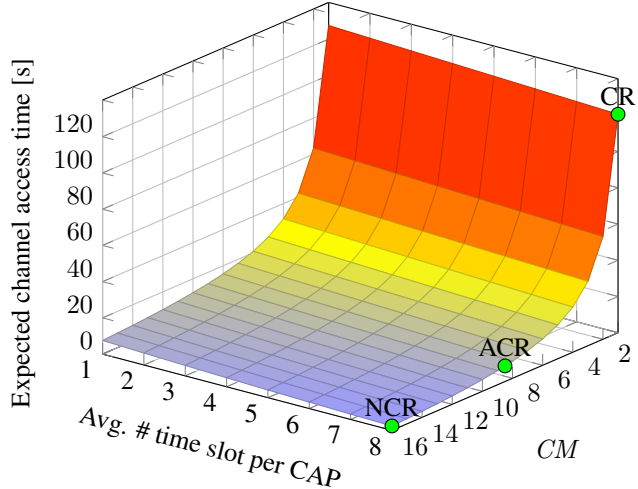

The expected channel excess time for different values of 
$\gls{NCAP}$ and $\gls{SCAP}$ is shown in 
Fig.~\ref{fig:channel_access_time}. Here, one can see that 
$\ACAP$ achieves a much lower expected channel access time 
than \CAPR\ mode. Therefore, the 
adaptability with \ACAP is higher. It can be seen that the 
expected channel access time scales exponentially with a 
decreasing CAP frequency and \ACAP operates at the part where 
a good adaptability is still given. In contrast to that, \DCAP 
reduces the number of slots in individual CAPs. Therefore, the 
expected time to send a CAP message is always quite low. However, 
in extreme cases all CAP slots might be allocated and \DCAP 
performs as \CAPR\ in terms of expected channel access time.
 
\section{Scenario Description} 
\label{sec:experimental_setup}
We focus on the widely used data-collection scenario using the
converge-cast pattern. In this pattern, routing is performed along a
rooted oriented tree. In particular we use a rooted binary tree
topology with 31 nodes. We abstain from using an unbalanced tree to
eliminate the influence of heavily skewed traffic. The chosen
topology allows to analyze the behavior of nodes with different loads,
because traffic load increases exponentially from the leaves to the
sink. All nodes except the sink generate an average of
$\gls{delta}=\{1,\dots,4\}$ packets per second and forward them along
the routing tree to the sink. The packet generation follows a Poisson
distribution with mean $\lambda$ and a one second observation interval
to model the dynamic behavior of nodes. Two packet generation patterns
are considered:
\begin{enumerate}
\item Generation of packets with $\lambda=\gls{delta}$ 
packets.
\item Generation of packets in bursts of size \gls{delta} and
  $\lambda=1$ burst. 
\end{enumerate}
The first scenario allows the assessment of the network close to
saturation because traffic is pretty stable. In the second scenario
(bursty traffic), the proposed CAP reduction mechanisms have to
(de)allocate GTSs continuously and adjust to the current traffic demand
of the network.

As discussed in Sect.~\ref{sec:metrics} and
Sect.~\ref{sec:theoretical_evaluation}, all CAP reduction mechanisms
only depend on the difference $\gls{MO}-\gls{SO}$. Therefore, either
\gls{SO} or \gls{MO} can remain constant for simulation. We set
$\gls{SO}$ to $3$ so that a single packet with 127 bytes can be sent
per GTS. Furthermore, it insures that all CAPs have the same length
for all configurations. If not stated otherwise, we also fixed
$\gls{BO}=7$ to guarantee sufficiently many beacon slots in the
network for all nodes to associate. The scenario is evaluated for
$4 \le \gls{MO} \le 7$. Each node's queue is divided in a CAP queue of
length $Q_{CAP}$, and a GTS queue of length $Q_{GTS}$, with a combined
total capacity of 30 packets. Tbl~\ref{tab:parameterValues} summarizes
the setup of the simulation. For each configuration, 20 runs are
conducted and results are shown with a 95\% confidence level.
Simulations are done using OMNeT++ and openDSME.

\begin{table}[h]
	\centering
	\caption{Setup of DSME parameters, traffic generator and TPS
          scheduling parameters.}
	\begin{tabular}{p{.12\textwidth}  p{.06\textwidth} p{.12\textwidth} p{.06\textwidth} p{.05\textwidth}  p{.09\textwidth} p{.09\textwidth} p{.12\textwidth} p{.25\textwidth}}    
		\hline
		\textbf{Parameter} & \gls{SO} & \gls{MO} & \gls{BO} & \gls{alpha} &  $Q_{CAP}$ & $Q_{GTS}$  &  \gls{delta} \\
		\hline
		\textbf{Values}& 3 & \{4, \dots, 7\}  & 7 & 0.1  & 8 & 22 & \{1, \dots, 4 \}   \\
		\hline\\[-1mm]
	\end{tabular}
	\label{tab:parameterValues}    
\end{table}


\section{Simulative evaluation} \label{sec:simulative_evaluation} In
the following, we present a performance assessment of the proposed
mechanisms by comparing them with \NCAPR and \CAPR. Performance
metrics include the packet reception ratio (PRR), the mean queue
length and the maximum number of GTSs allocated. Additionally, the
adaptability to time varying traffic metric as explicated in
Sect.~\ref{sec:theoretical_evaluation} is also evaluated by
simulation. It corresponds to the GTS-negotiation message dwell time.
Dwell time is the time between the generation of a message and its
transmission, including queuing delay. Therefore, it provides an
insight into a system's adaptability because the system adapts to
changes in traffic faster by (de)allocating GTSs if the respective
control messages have a lower dwell time.

\subsection{Varying burst sizes} 
\label{sec:varying_generation_intervals}
\begin{figure}[]
	\centering
	\input{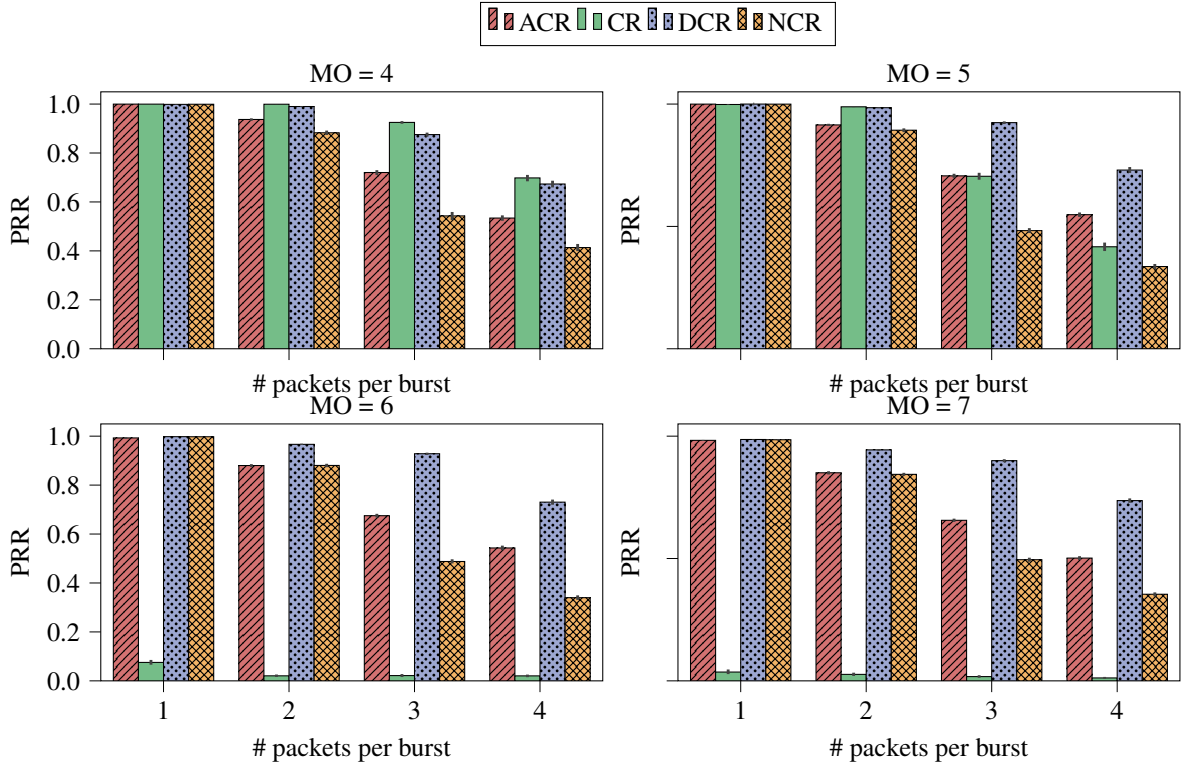}
	\caption{PRR for varying numbers of packets per burst and
		different values of \gls{MO}.}
	\label{fig:dcap_burst_mo}  
\end{figure}

Fig.~\ref{fig:dcap_burst_mo} shows the average packet reception ratio
(PRR) for an increasing \gls{delta} and different values of \gls{MO}
with a $\lambda=1$ burst. For all methods, the PRR decreases for an
increasing number of packets per burst. This is mainly due to the
following reason: especially for smaller values of \gls{MO}, there are
not enough GTSs per second to accommodate all packets generated during
a burst, leading to dropped packets as the queues fill up.

For \CAPR, the frequency of CAPs decreases for increasing \gls{MO},
resulting in more contention during the remaining CAPs. Thus, the
required GTSs cannot be allocated in time. This is the case for
$\gls{MO}=5$, where $\gls{tau}$ equals $81.25\%$. From this
theoretical value, a high PRR was expected. However our results show
that for an increasing number of packets per burst \CAPR is very
sensible to high traffic loads.

The performance of \NCAPR is independent of \gls{MO}, as already
indicated by the theoretical analysis in
Sect.~\ref{sec:theoretical_evaluation}. Therefore, the choice of \CAPR
or \NCAPR depends largely on \gls{MO}, as \CAPR performs better than
\NCAPR for $\gls{MO}=4$ and $\gls{MO}=5$, but \NCAPR performs better
for $\gls{MO}=6$ and $\gls{MO}=7$. If $\gls{SO}=\gls{MO}=3$, all
mechanisms perform equally well, since there is only a single \SF per
\MSF, which cannot be reduced by \DCAP or \CAPR.

On the other hand, \DCAP performs similarly to \CAPR for $\gls{MO}=4$
and outperforms \CAPR and \NCAPR for $\gls{MO} \ge 5$. That is because
\DCAP starts with \NCAPR mode and therefore allocates GTSs as fast as
possible (e.g., as fast as \NCAPR) for larger values of \gls{MO} where
there is only a small number of CAPs for \CAPR. In addition, it
reduces CAPs when all regular GTSs are already allocated so that it
converges towards \CAPR. However, the allocation of these
additional resources has a slight time overhead, resulting in the
performance gap between \DCAP and \CAPR for $\gls{MO}=4$.

Similarly to \CAPR, the performance of \ACAP depends on \gls{MO} but
its influence is not as strong as for \CAPR. For smaller values of \gls{MO}, \ACAP's PRR
is about the average between values for \CAPR and \NCAPR. It
corresponds to the theoretical analysis regarding \gls{tau}, with
boundaries delimited by \CAPR and \NCAPR (e.g. $\gls{tau} = 56.25\%$
for $\gls{MO}= 4$). For larger values of \gls{MO} (i.e.
$\gls{MO} \ge 6$), the PRR is less sensible to this parameter,
performing even better than \NCAPR. This is because of the alternating
behavior every \BI, which provides enough CAPs to perform the required
GTS negotiations. Moreover, the fact that \ACAP allocates first
\GTSCFP and then less frequent GTSs (i.e. \GTSCAP) is a key aspect in
the performance of this mechanism.

All in all, \ACAP and \DCAP are attractive alternatives to DSME under
varying traffic patterns. The former because it combines the strengths
of \CAPR and \NCAPR mechanisms in one approach. The latter because it
dynamically adapts to the traffic demands of the network and
(de)allocates additional GTSs during CAPs to increase reliability.

\subsection{Varying packet generation rates}
\begin{figure}[]
	\centering
	\input{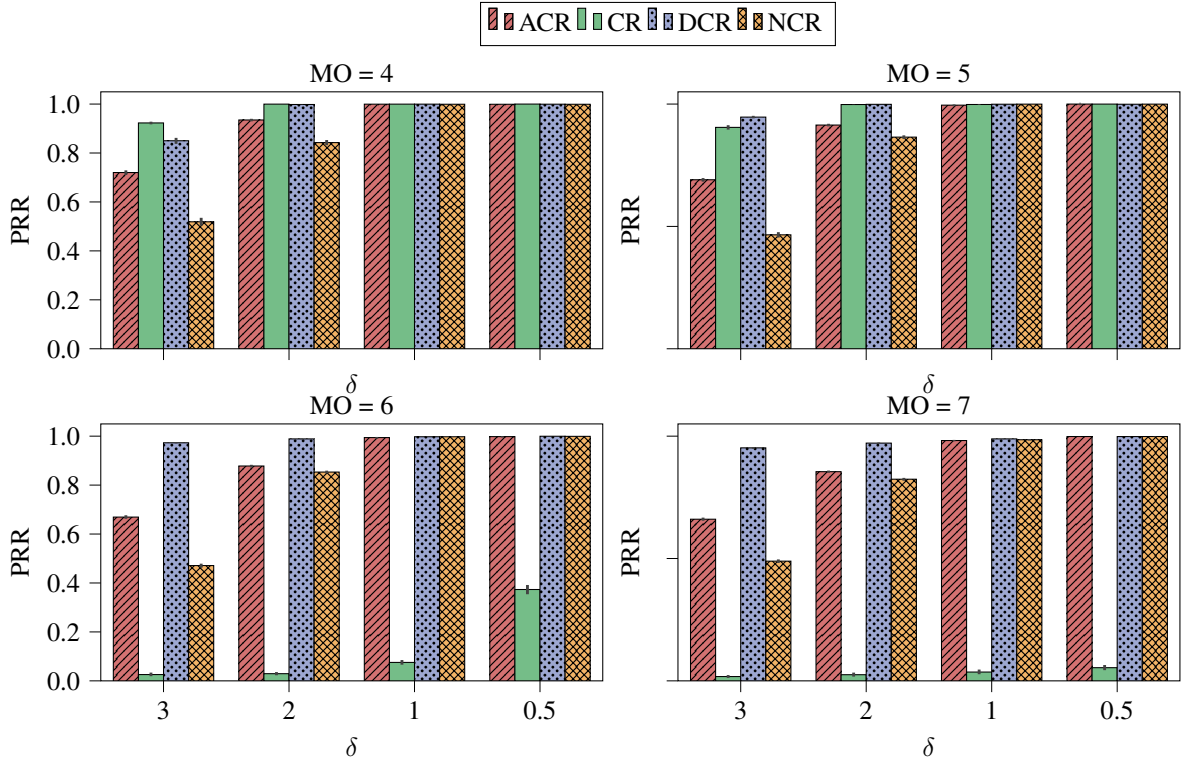}
	\caption{PRR of \DCAP, \ACAP, \CAPR and \NCAPR for different values 
		of \gls{MO} and $\delta$.}
	\label{fig:dcap_exp_prr}
\end{figure}

The following figures show a performance assessment for different
packet generation intervals, which correspond to $1 / \delta$ seconds.
Particularly, Fig.~\ref{fig:dcap_exp_prr} depicts the PRR for
different values of \gls{MO}. As already explained in
Sect.~\ref{sec:varying_generation_intervals}, the choice of \CAPR and
\NCAPR largely depends on \gls{MO}, with \CAPR performing better for
$\gls{MO} \le 5$ and \NCAPR performing better for $\gls{MO} \ge 6$.
\NCAPR, \ACAP and \DCAP provide a good performance for
$\gls{delta} \le 1$ and all values of \gls{MO}. For $\gls{delta} > 1$
\NCAPR is operating beyond the maximum capacity and \ACAP starts to
reach the limits of the capacity of available GTSs. Therefore,
performance of \NCAPR and \ACAP is diminished. Contrary to \NCAPR,
\ACAP's PRR is also slightly affected by increasing \gls{MO}. \DCAP performs significantly better than the other CAP reduction mechanisms, especially for larger values of \gls{MO}. E.g., it achieves a PRR of 95\% for $\gls{MO}=7$ and $\gls{delta}=3$, while \NCAPR achieves only about 48\%, \ACAP about 67\% and \CAPR less than 2\%.

\begin{figure}[]
    \centering
    \input{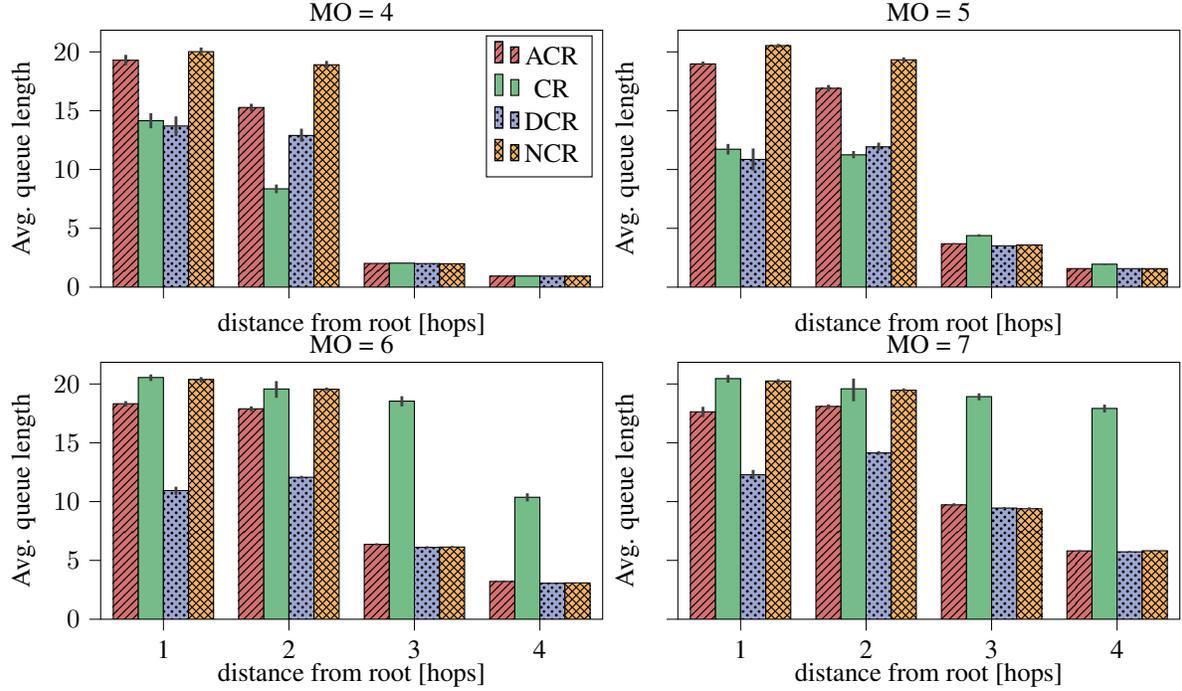}
    \caption{Average queue length per node for different values 
    of \gls{MO} and $\gls{delta}=3$. Nodes with the same 
    number of hops to the root 
    are grouped together.}
    \label{fig:dcap_exp_q}
\end{figure}

The main reason for lost packets are queue drops. This is also
reflected by the average queue length, as shown in
Fig.~\ref{fig:dcap_exp_q}. Here, nodes with the same distance to the
sink are grouped together, as nodes closer to the sink experience more
traffic than nodes further away. With a maximum of 22 packets for
$Q_{GTS}$, it is clearly visible that \NCAPR operates close to the
maximum queue capacity at nodes close to the sink for all values of
\gls{MO}. That is because the network is over-saturated for a packet
generation interval of 0.33 seconds and no more GTSs are available.
For \ACAP the network is close to the maximum capacity and therefore
the average queue length for nodes closer to the sink is significantly
higher and close to the maximum queue capacity (i.e. nodes up to 2
hops away from the sink node). This funneling effect is intensified
for increasing \gls{MO}, which reduces the frequency of CAPs per \MSF
in \textit{BI}s in which \ACAP operates in \CAPR mode. On the other
hand, the average queue length for \CAPR increases with increasing
\gls{MO} because there are fewer CAPs per \MSF. Consequently, the
remaining CAPs are more congested and the required GTSs are not
allocated in time. The average queue length for \DCAP is lower than
68\% of the maximum queue capacity for all values of \gls{MO}.


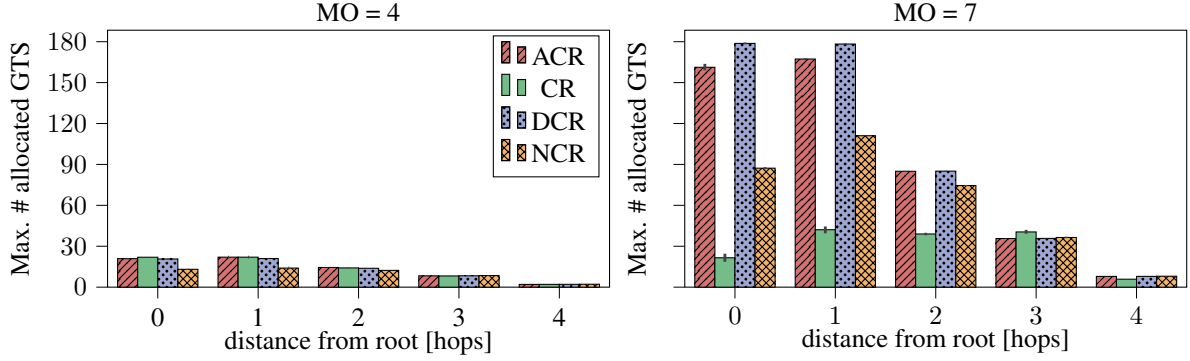
\begin{figure}[]
    \centering
\begin{tikzpicture}
\pgfplotsset{compat=newest}
\definecolor{color0}{RGB}{212, 114, 114}
\definecolor{color1}{RGB}{117, 189, 136}
\definecolor{color2}{RGB}{157, 167, 209}
\definecolor{color3}{RGB}{242, 179, 102}

\begin{groupplot}[group style={group size=2 by 1}]
\nextgroupplot[
height=5cm,
scaled x ticks=manual:{}{\pgfmathparse{#1}},
tick align=outside,
tick pos=left,
title={MO = 4},
width=.5\textwidth,
x grid style={white!69.0196078431373!black},
xlabel={distance from root [hops]},
xmin=-0.5, xmax=4.5,
xtick style={color=black},
xticklabels={0,1,2,3,4},
y grid style={white!69.0196078431373!black},
ylabel={Max. \# allocated GTS},
ymin=0, ymax=188.4225,
ytick style={color=black},
ytick = {0, 30, 60, 90, 120, 150, 180},
x label style={yshift=.5em},
title style={yshift=-.6em},
xtick={0,1,2,3,4},
]
\draw[draw=black,fill=color0, postaction={pattern=north 
	east lines}] (axis cs:-0.4,0) rectangle (axis 
	cs:-0.2,20.9);
\addlegendimage{ybar,ybar 
legend,draw=black,fill=color0, 
postaction={pattern=north 
	east lines}};
\addlegendentry{\ACAP}

\draw[draw=black,fill=color0, postaction={pattern=north 
	east lines}] (axis cs:0.6,0) rectangle (axis cs:0.8,22);
\draw[draw=black,fill=color0, postaction={pattern=north 
	east lines}] (axis cs:1.6,0) rectangle (axis 
	cs:1.8,14.45);
\draw[draw=black,fill=color0, postaction={pattern=north 
	east lines}] (axis cs:2.6,0) rectangle (axis 
	cs:2.8,8.3125);
\draw[draw=black,fill=color0, postaction={pattern=north 
	east lines}] (axis cs:3.6,0) rectangle (axis 
	cs:3.8,2.025);
\draw[draw=black,fill=color1] (axis cs:-0.2,0) rectangle 
(axis cs:0,21.95);
\addlegendimage{ybar,ybar 
legend,draw=black,fill=color1};
\addlegendentry{\CAPR}

\draw[draw=black,fill=color1] (axis cs:0.8,0) rectangle 
(axis cs:1,22);
\draw[draw=black,fill=color1] (axis cs:1.8,0) rectangle 
(axis cs:2,14.1125);
\draw[draw=black,fill=color1] (axis cs:2.8,0) rectangle 
(axis cs:3,8.21875);
\draw[draw=black,fill=color1] (axis cs:3.8,0) rectangle 
(axis cs:4,2.053125);
\draw[draw=black,fill=color2, 
postaction={pattern=crosshatch dots}] (axis 
cs:2.77555756156289e-17,0) rectangle (axis cs:0.2,20.65);
\addlegendimage{ybar,ybar 
legend,draw=black,fill=color2, 
postaction={pattern=crosshatch dots}};
\addlegendentry{\DCAP}

\draw[draw=black,fill=color2, 
postaction={pattern=crosshatch dots}] (axis cs:1,0) 
rectangle (axis cs:1.2,21);
\draw[draw=black,fill=color2, 
postaction={pattern=crosshatch dots}] (axis cs:2,0) 
rectangle (axis cs:2.2,13.8375);
\draw[draw=black,fill=color2, 
postaction={pattern=crosshatch dots}] (axis cs:3,0) 
rectangle (axis cs:3.2,8.31875);
\draw[draw=black,fill=color2, 
postaction={pattern=crosshatch dots}] (axis cs:4,0) 
rectangle (axis cs:4.2,2.084375);
\draw[draw=black,fill=color3, 
postaction={pattern=crosshatch}] (axis cs:0.2,0) 
rectangle (axis cs:0.4,13.1);
\addlegendimage{ybar,ybar 
legend,draw=black,fill=color3, 
postaction={pattern=crosshatch}};
\addlegendentry{\NCAPR}

\draw[draw=black,fill=color3, 
postaction={pattern=crosshatch}] (axis cs:1.2,0) 
rectangle (axis cs:1.4,14);
\draw[draw=black,fill=color3, 
postaction={pattern=crosshatch}] (axis cs:2.2,0) 
rectangle (axis cs:2.4,12.3);
\draw[draw=black,fill=color3, 
postaction={pattern=crosshatch}] (axis cs:3.2,0) 
rectangle (axis cs:3.4,8.50625);
\draw[draw=black,fill=color3, 
postaction={pattern=crosshatch}] (axis cs:4.2,0) 
rectangle (axis cs:4.4,2.15);
\addplot [line width=1.08pt, white!26!black]
table {%
-0.3 20.75
-0.3 21
};
\addplot [line width=1.08pt, white!26!black]
table {%
0.7 22
0.7 22
};
\addplot [line width=1.08pt, white!26!black]
table {%
1.7 14.325
1.7 14.575
};
\addplot [line width=1.08pt, white!26!black]
table {%
2.7 8.15625
2.7 8.53125
};
\addplot [line width=1.08pt, white!26!black]
table {%
3.7 2.0125
3.7 2.0375
};
\addplot [line width=1.08pt, white!26!black]
table {%
-0.1 21.85
-0.1 22
};
\addplot [line width=1.08pt, white!26!black]
table {%
0.9 22
0.9 22
};
\addplot [line width=1.08pt, white!26!black]
table {%
1.9 14.025
1.9 14.2
};
\addplot [line width=1.08pt, white!26!black]
table {%
2.9 8.125
2.9 8.325
};
\addplot [line width=1.08pt, white!26!black]
table {%
3.9 2.025
3.9 2.084375
};
\addplot [line width=1.08pt, white!26!black]
table {%
0.1 20.35
0.1 20.95
};
\addplot [line width=1.08pt, white!26!black]
table {%
1.1 20.825
1.1 21.175
};
\addplot [line width=1.08pt, white!26!black]
table {%
2.1 13.7125
2.1 13.975
};
\addplot [line width=1.08pt, white!26!black]
table {%
3.1 8.21875
3.1 8.425
};
\addplot [line width=1.08pt, white!26!black]
table {%
4.1 2.05625
4.1 2.11875
};
\addplot [line width=1.08pt, white!26!black]
table {%
0.3 12.75
0.3 13.45
};
\addplot [line width=1.08pt, white!26!black]
table {%
1.3 14
1.3 14
};
\addplot [line width=1.08pt, white!26!black]
table {%
2.3 12.1625
2.3 12.4375
};
\addplot [line width=1.08pt, white!26!black]
table {%
3.3 8.31875
3.3 8.7125
};
\addplot [line width=1.08pt, white!26!black]
table {%
4.3 2.10625
4.3 2.190625
};

\nextgroupplot[
height=5cm,
scaled y ticks=manual:{}{\pgfmathparse{#1}},
tick align=outside,
tick pos=left,
title={MO = 7},
width=.5\textwidth,
x grid style={white!69.0196078431373!black},
xlabel={distance from root [hops]},
xmin=-0.5, xmax=4.5,
xtick={0,1,2,3,4},
xtick style={color=black},
y grid style={white!69.0196078431373!black},
ylabel={Max. \# allocated GTS},
ymin=0, ymax=188.4225,
ytick style={color=black},
ytick = {0, 30, 60, 90, 120, 150, 180},
yticklabels={},
x label style={yshift=.5em},
title style={yshift=-.6em},
]
\draw[draw=black,fill=color0, postaction={pattern=north 
	east lines}] (axis cs:-0.4,0) rectangle (axis 
	cs:-0.2,161.25);
\draw[draw=black,fill=color0, postaction={pattern=north 
	east lines}] (axis cs:0.6,0) rectangle (axis 
	cs:0.8,167.3);
\draw[draw=black,fill=color0, postaction={pattern=north 
	east lines}] (axis cs:1.6,0) rectangle (axis 
	cs:1.8,84.9875);
\draw[draw=black,fill=color0, postaction={pattern=north 
	east lines}] (axis cs:2.6,0) rectangle (axis 
	cs:2.8,35.65625);
\draw[draw=black,fill=color0, postaction={pattern=north 
	east lines}] (axis cs:3.6,0) rectangle (axis 
	cs:3.8,7.88125);
\draw[draw=black,fill=color1] (axis cs:-0.2,0) rectangle 
(axis cs:0,21.55);
\draw[draw=black,fill=color1] (axis cs:0.8,0) rectangle 
(axis cs:1,42.1);
\draw[draw=black,fill=color1] (axis cs:1.8,0) rectangle 
(axis cs:2,39);
\draw[draw=black,fill=color1] (axis cs:2.8,0) rectangle 
(axis cs:3,40.49375);
\draw[draw=black,fill=color1] (axis cs:3.8,0) rectangle 
(axis cs:4,5.834375);
\draw[draw=black,fill=color2, 
postaction={pattern=crosshatch dots}] (axis 
cs:2.77555756156289e-17,0) rectangle (axis 
cs:0.2,178.75);
\draw[draw=black,fill=color2, 
postaction={pattern=crosshatch dots}] (axis cs:1,0) 
rectangle (axis cs:1.2,178.3);
\draw[draw=black,fill=color2, 
postaction={pattern=crosshatch dots}] (axis cs:2,0) 
rectangle (axis cs:2.2,85.075);
\draw[draw=black,fill=color2, 
postaction={pattern=crosshatch dots}] (axis cs:3,0) 
rectangle (axis cs:3.2,35.75625);
\draw[draw=black,fill=color2, 
postaction={pattern=crosshatch dots}] (axis cs:4,0) 
rectangle (axis cs:4.2,7.909375);
\draw[draw=black,fill=color3, 
postaction={pattern=crosshatch}] (axis cs:0.2,0) 
rectangle (axis cs:0.4,87.25);
\draw[draw=black,fill=color3, 
postaction={pattern=crosshatch}] (axis cs:1.2,0) 
rectangle (axis cs:1.4,111.05);
\draw[draw=black,fill=color3, 
postaction={pattern=crosshatch}] (axis cs:2.2,0) 
rectangle (axis cs:2.4,74.5375);
\draw[draw=black,fill=color3, 
postaction={pattern=crosshatch}] (axis cs:3.2,0) 
rectangle (axis cs:3.4,36.43125);
\draw[draw=black,fill=color3, 
postaction={pattern=crosshatch}] (axis cs:4.2,0) 
rectangle (axis cs:4.4,8.015625);
\addplot [line width=1.08pt, white!26!black]
table {%
-0.3 159.05
-0.3 163.75125
};
\addplot [line width=1.08pt, white!26!black]
table {%
0.7 166.675
0.7 167.925625
};
\addplot [line width=1.08pt, white!26!black]
table {%
1.7 84.625
1.7 85.325
};
\addplot [line width=1.08pt, white!26!black]
table {%
2.7 35.45625
2.7 35.86265625
};
\addplot [line width=1.08pt, white!26!black]
table {%
3.7 7.815625
3.7 7.965625
};
\addplot [line width=1.08pt, white!26!black]
table {%
-0.1 18.6
-0.1 24.5
};
\addplot [line width=1.08pt, white!26!black]
table {%
0.9 39.475
0.9 44.5
};
\addplot [line width=1.08pt, white!26!black]
table {%
1.9 38.0371875
1.9 39.8753125
};
\addplot [line width=1.08pt, white!26!black]
table {%
2.9 39.19984375
2.9 41.99375
};
\addplot [line width=1.08pt, white!26!black]
table {%
3.9 5.709296875
3.9 5.968828125
};
\addplot [line width=1.08pt, white!26!black]
table {%
0.1 178.05
0.1 179.45
};
\addplot [line width=1.08pt, white!26!black]
table {%
1.1 177.9
1.1 178.700625
};
\addplot [line width=1.08pt, white!26!black]
table {%
2.1 84.8246875
2.1 85.3125
};
\addplot [line width=1.08pt, white!26!black]
table {%
3.1 35.55625
3.1 35.95
};
\addplot [line width=1.08pt, white!26!black]
table {%
4.1 7.85625
4.1 7.959375
};
\addplot [line width=1.08pt, white!26!black]
table {%
0.3 86.45
0.3 88.00125
};
\addplot [line width=1.08pt, white!26!black]
table {%
1.3 110.85
1.3 111.275
};
\addplot [line width=1.08pt, white!26!black]
table {%
2.3 74.1625
2.3 74.9125
};
\addplot [line width=1.08pt, white!26!black]
table {%
3.3 35.8996875
3.3 37.00640625
};
\addplot [line width=1.08pt, white!26!black]
table {%
4.3 7.9375
4.3 8.106328125
};
\end{groupplot}

\end{tikzpicture}
    \caption{Maximum number of allocated GTSs per node 
    for $\gls{MO}=4$, $\gls{MO}=7$ and 
    $\gls{delta}=3$. Nodes with the same number of hops 
    to the root are grouped together.}
    \label{fig:dcap_exp_slots}
\end{figure}

The maximum number of RX- and TX-GTSs allocated by the different \CAPR
mechanisms for $\gls{delta}=3$ and different values of \gls{MO} is
illustrated in Fig.~\ref{fig:dcap_exp_slots}. Nodes with the same
distance to the sink are grouped together, as more GTSs are allocated
closer to the sink. For $\gls{MO}=4$, less slots need to be allocated
because the \MSF - and thus the schedule of allocated GTSs - is
repeated multiple times per packet generation interval. Theoretically,
for $\gls{MO}=7$ and $\gls{delta}= 3$, the number of required GTSs at
the sink node to successfully receive all packets generated in the
network is about $180$ GTSs per \BI (i.e. assuming that no packet was
dropped from any queue). For \NCAPR and \CAPR the highest number of
available GTSs per \BI amounts to $112$ and $232$ GTSs respectively.
For \ACAP this value corresponds to the average between \NCAPR and
\CAPR, which is about $172$ GTSs per \BI. However, given the funneling
effect that is inherent for data collection scenarios this
assumption does not hold. That is, many packets are lost in nodes
closer the sink node and therefore the number of allocated GTSs are a
bit less than the theoretical value. That is the case for \NCAPR, in
which nodes 1 hop further require about $174$ GTSs. However, only
$112$ GTSs can be negotiated. Therefore, about $90$ GTSs are allocated
at the sink node, which supports the obtained PRR with a value of
$47\%$.

For \ACAP, one could expect a good performance of the network with
these configuration parameters since the average number of available
GTSs (i.e. $172$ GTSs) should be sufficient to guarantee a low number
of packet drops. However, this is not the case because the number of
usable GTS varies over time. That is, the number of usable GTSs is
limited to $112$ in one \BI~ -- while operating in \NCAPR~ -- and
lifted to a maximum of $232$ GTSs in the next \BI~ -- while
operating in \CAPR. The effect of this drastic reduction of GTSs from
one \BI to the other is observed in the large value of the average
queue lengths for nodes close to the sink. That is, despite the fact
that all usable GTSs are allocated, the higher number of queued
packets in the \BI operating in \NCAPR leads to queue overflows,
specially for nodes closer to the sink node.

For \DCAP, the largest number of GTS is allocated. Theoretically,
\CAPR could reach the same performance as \DCAP but it fails to
allocate the GTS in time, as already explained. Still, \DCAP performs
well for $\gls{MO}\le 7$ because it can allocate a large number of GTS
at the start and then it successively reduces the CAPs to allocate the
last remaining GTSs.

\begin{figure}[]
	\centering
\usetikzlibrary{patterns}

\begin{tikzpicture}
\definecolor{color0}{RGB}{212, 114, 114}
\definecolor{color1}{RGB}{117, 189, 136}
\definecolor{color2}{RGB}{157, 167, 209}
\definecolor{color3}{RGB}{242, 179, 102}

\begin{axis}[
legend columns=1,
height=5cm,
legend cell align={left},
legend style={fill opacity=0.8, draw opacity=1, text 
opacity=1, at={(1.4,0.5)}, anchor=east, 
draw=white!80!black},
log basis y={10},
tick align=outside,
tick pos=left,
width=.6\textwidth,
x grid style={white!69.0196078431373!black},
xlabel={MO},
xmin=-0.5, xmax=3.5,
xtick style={color=black},
xtick={0,1,2,3},
xticklabels={4,5,6,7},
y grid style={white!69.0196078431373!black}, 
ylabel={dwell time [s]},
ymin=0.0127665673660517, ymax=2.92030479252542,
ymode=log,
ytick style={color=black},
ytick={0.001,0.01,0.1,1,10,100},
yticklabels={\(\displaystyle {10^{-3}}\),\(\displaystyle {10^{-2}}\),\(\displaystyle {10^{-1}}\),\(\displaystyle {10^{0}}\),\(\displaystyle {10^{1}}\),\(\displaystyle {10^{2}}\)}
]
\draw[draw=black,fill=color0, postaction={pattern=north 
	east lines}] (axis cs:-0.4,0.0000001) 
rectangle 
(axis cs:-0.2,0.0427990813974704);
\addlegendimage{ybar,ybar 
legend,draw=black,fill=color0, postaction={pattern=north 
		east lines}};
\addlegendentry{\ACAP}

\draw[draw=black,fill=color0, postaction={pattern=north 
east lines}] (axis 
cs:0.6,0.0000001) 
rectangle 
(axis cs:0.8,0.105038016057531);
\draw[draw=black,fill=color0, postaction={pattern=north 
	east lines}] (axis cs:1.6,0.0000001) 
rectangle (axis cs:1.8,0.208195920249805);
\draw[draw=black,fill=color0, postaction={pattern=north 
	east lines}] (axis cs:2.6,0.0000001) 
rectangle (axis cs:2.8,0.413479946612835);
\draw[draw=black,fill=color1, postaction={}] 
(axis cs:-0.2,0.0000001) 
rectangle (axis cs:0,0.0580091204560533);
\addlegendimage{ybar,ybar 
legend,draw=black,fill=color1, postaction={}};
\addlegendentry{\CAPR}

\draw[draw=black,fill=color1, postaction={}] (axis 
cs:0.8,0.0000001) 
rectangle (axis cs:1,0.180650952975979);
\draw[draw=black,fill=color1, postaction={}] (axis 
cs:1.8,0.0000001) 
rectangle (axis cs:2,0.769310745042989);
\draw[draw=black,fill=color1, postaction={}] (axis 
cs:2.8,0.0000001) 
rectangle (axis cs:3,2.21245892412208);
\draw[draw=black,fill=color2, 
postaction={pattern=crosshatch dots}] 
(axis 
cs:2.77555756156289e-17,0.0000001) rectangle (axis 
cs:0.2,0.0170814622423894);
\addlegendimage{ybar,ybar 
legend,draw=black,fill=color2, 
postaction={pattern=crosshatch dots}};
\addlegendentry{\DCAP}

\draw[draw=black,fill=color2, 
postaction={pattern=crosshatch dots}] 
(axis cs:1,0.0000001) 
rectangle (axis cs:1.2,0.0181941196604819);
\draw[draw=black,fill=color2, 
postaction={pattern=crosshatch dots}] 
(axis cs:2,0.0000001) 
rectangle (axis cs:2.2,0.0186007008918685);
\draw[draw=black,fill=color2, 
postaction={pattern=crosshatch dots}] 
(axis cs:3,0.0000001) 
rectangle (axis cs:3.2,0.0226096859435627);
\draw[draw=black,fill=color3, 
postaction={pattern=crosshatch}] 
(axis cs:0.2,0.0000001) 
rectangle (axis cs:0.4,0.0180609649819566);
\addlegendimage{ybar,ybar 
legend,draw=black,fill=color3, 
postaction={pattern=crosshatch}};
\addlegendentry{\NCAPR}

\draw[draw=black,fill=color3, 
postaction={pattern=crosshatch}] (axis 
cs:1.2,0.0000001) 
rectangle (axis cs:1.4,0.0223109577279569);
\draw[draw=black,fill=color3, 
postaction={pattern=crosshatch}] (axis 
cs:2.2,0.0000001) 
rectangle (axis cs:2.4,0.0245427848884168);
\draw[draw=black,fill=color3, 
postaction={pattern=crosshatch}] (axis 
cs:3.2,0.0000001) 
rectangle (axis cs:3.4,0.029221075423745);
\addplot [line width=1.08pt, white!26!black, forget plot]
table {%
-0.3 0.0410146680139408
-0.3 0.0444817137155161
};
\addplot [line width=1.08pt, white!26!black, forget plot]
table {%
0.7 0.101555589644759
0.7 0.108675145064837
};
\addplot [line width=1.08pt, white!26!black, forget plot]
table {%
1.7 0.199762999990409
1.7 0.217041566197328
};
\addplot [line width=1.08pt, white!26!black, forget plot]
table {%
2.7 0.398027656457228
2.7 0.431017147036657
};
\addplot [line width=1.08pt, white!26!black, forget plot]
table {%
-0.1 0.0559503872688144
-0.1 0.0600272359217165
};
\addplot [line width=1.08pt, white!26!black, forget plot]
table {%
0.9 0.175995986517941
0.9 0.185064741894978
};
\addplot [line width=1.08pt, white!26!black, forget plot]
table {%
1.9 0.74562241306971
1.9 0.792872878126766
};
\addplot [line width=1.08pt, white!26!black, forget plot]
table {%
2.9 2.14200704515407
2.9 2.28131268704474
};
\addplot [line width=1.08pt, white!26!black, forget plot]
table {%
0.1 0.0163424628613606
0.1 0.0178328755293271
};
\addplot [line width=1.08pt, white!26!black, forget plot]
table {%
1.1 0.0174978037862102
1.1 0.0188045175145142
};
\addplot [line width=1.08pt, white!26!black, forget plot]
table {%
2.1 0.0179321280487459
2.1 0.0192813325768217
};
\addplot [line width=1.08pt, white!26!black, forget plot]
table {%
3.1 0.0219640110892356
3.1 0.0232584395695798
};
\addplot [line width=1.08pt, white!26!black, forget plot]
table {%
0.3 0.0173050710631392
0.3 0.0187427280103664
};
\addplot [line width=1.08pt, white!26!black, forget plot]
table {%
1.3 0.0216702972137915
1.3 0.0229292675853149
};
\addplot [line width=1.08pt, white!26!black, forget plot]
table {%
2.3 0.0240625621925042
2.3 0.0250348836099172
};
\addplot [line width=1.08pt, white!26!black, forget plot]
table {%
3.3 0.0287901281711764
3.3 0.0296293528405195
};
\end{axis}

\end{tikzpicture}
	\caption{Average dwell time for GTS-negotiation messages 
		using \CAPR, \NCAPR, \ACAP and \DCAP for different values of 
		\gls{MO} and a packet generation rate of 1 packet per 
		second.}
	\label{fig:dcap_burst_channel_access_time}
\end{figure}
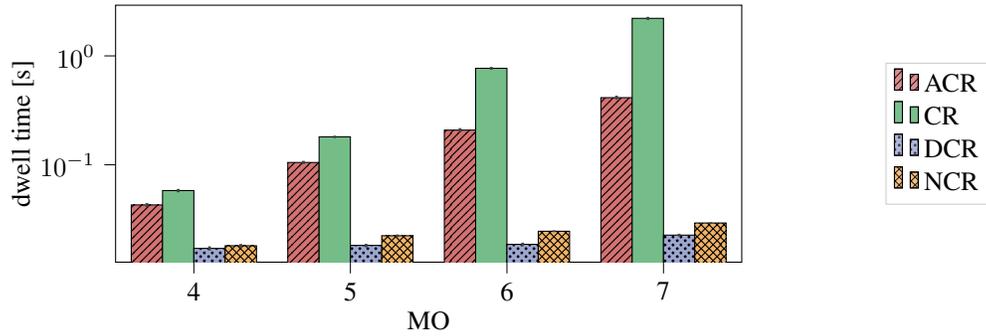

Evaluation of the average dwell time, as illustrated in Fig.~\ref{fig:dcap_burst_channel_access_time}, with one packet per second for
messages sent during the GTS negotiation, i.e., \GTSRQ, \GTSRP and \GTSNF
shows that \CAPR has the highest dwell time. That is because there is
only a single CAP per \MSF, and GTS commands have to be queued for a
long time. The dwell time exponentially increases for an increasing
\gls{MO} as the duration of a MSF doubles when incrementing \gls{MO}.
\ACAP comes second in this ranking, which has a similar behavior as
\CAPR. Although \ACAP's dwell time is influenced by the length of the
MSF given by \gls{MO}, this effect is diminished during the \gls{BI}s
in which \ACAP operates in \NCAPR. It should be noted, that although
\ACAP's waiting time is higher than \DCAP and \NCAPR, it is still
enough to guarantee adaptability to fluctuating traffic. The
validation of \ACAP's dwell time by simulations confirm the
theoretically estimated value (Sec.~\ref{sec:theoretical_evaluation}).
This is the case for $\gls{MO} = 4$ with an average dwell time of
42.80 ms against 44.64 ms from the theoretical model. In this case,
the difference between the values comes from the way the estimation
in the theoretical model is performed (i.e. average between \NCAPR
and \CAPR). That is, the theoretical model does not take into account
the scheduling mechanism implemented for \ACAP, which reduces on a
small scale its dwell time.

\DCAP and \NCAPR perform equally well, e.g., with an average dwell
time of 17.08 ms and 18.06 ms, respectively. For \NCAPR this matches
with the value of about 17.28 ms from the theoretical models
(Sec.~\ref{sec:theoretical_evaluation} and
~\ref{sec:appendix:ch_time_ncapr}), which do not consider packet
collisions. \NCAPR has a higher dwell time than \DCAP because nodes
are unable to allocate enough GTSs (network saturation) but still can
send GTS commands for the required GTSs. This results in more
congestion during the CAPs and the nodes backing of more frequently
during the CSMA/CA algorithm. Both CAP reduction mechanisms are
independent of \gls{MO} with the dwell time only marginally increasing
for an increasing \gls{MO} because more GTSs are required initially
for larger values of \gls{MO} resulting in more contention. For
$\gls{MO}=3$ all mechanisms perform the same.

\subsection{Summary of results}
The following statements summarize the most significant results from
Sect.~\ref{sec:theoretical_evaluation} and
\ref{sec:simulative_evaluation} without claim to completeness. It is
assumed that the configuration parameters \gls{MO} and \gls{SO} are
fixed before the simulation according to the needs of the
application. 
We evaluated the capability of the proposed CAP reduction mechanisms
to support fluctuating traffic in comparison with the \NCAPR and \CAPR
modes provided by DSME. To this end, we tested two different packet
generation patterns: varying burst sizes and varying packet generation
rates with $\delta$ packets per second. Results are given from a data
collection scenario for a binary tree topology with 31 nodes.

\begin{itemize}
\item \textit{\NCAPR} provides good adaptability for all values of
  $\gls{MO}-\gls{SO}$. However, for a large number of packets per
  second ($\delta > 2$), \NCAPR cannot provide a sufficient number of
  GTSs for sending all packets. This results in a large number of
  dropped packets due to full queues.
\item \textit{\CAPR} performs well for $\gls{MO}-\gls{SO}\le2$. Here,
  a single CAP per MSF is sufficient to allocate all required GTSs and
  \CAPR still offers a higher throughput which increases the overall
  PRR. For $\gls{MO}-\gls{SO} \ge 3$, the performance of \CAPR
  diminishes since it is unable to allocate all GTS in time, resulting
  in large queues and packet loss.
\item \textit{\ACAP} offers a compromise between \CAPR and \NCAPR,
  which is reflected in its performance. For $\gls{MO}-\gls{SO}\le 1$,
  it performs as a middle ground between \CAPR and \NCAPR. For
  $\gls{MO}-\gls{SO}\ge 2$, \ACAP starts to outperform the standard
  modes of DSME because it offers a higher adaptability than \CAPR and
  a higher throughput than \NCAPR.
\item \textit{\DCAP} achieves the best overall performance in terms of
  PRR, dwell time, and queue utilization for all differences
  $\gls{MO}-\gls{SO}$ and different packet generation rates. That is
  because \DCAP starts in \NCAPR mode which offers a high
  adaptability. Then it gradually and dynamically reduces the CAP to
  allow for a higher throughput. Notably, it manages to achieve a PRR
  of over 80\% regardless of the value of $\gls{MO}-\gls{SO}$ in a
  scenario without bursts and therefore outperforms the standard modes
  of DSME.
\end{itemize}

%
\section{Conclusion}

One of the pillars of the Internet of Things (IoT) is a robust and
reliable wireless communication infrastructure. The IEEE 802.15.4 Deterministic and Synchronous Multi-Channel Extension (DSME) is a MAC protocol that guarantees reliability, scalability and energy-efficiency in WLANs. DSME offers a TDMA/FDMA-based channel access and provides the possibility to assign at run-time the resources time and frequency to individual links in a conflict free way. This allows a continuous distribution of the resources among the network nodes. This way the network can adopt itself to varying communication patterns. The redistribution of resources pursues two conflicting objectives, a high responsiveness towards bursts of packets on the one hand and an efficient utilization of the available bandwidth on the other hand. The first goal calls for a fast channel access for messages that (de)allocate resources.
In terms of DSME this calls for long and frequent CAP phases. In
contrast the second goal asks for frequent and long CFP phases, such
that application packets can be sent with minimal delay and in high
number. In DSME the balance between the goals agility and throughput
is controlled by \gls{tau}, the fraction  of CFP's time slots in a dataframe. A high value guarantees a high throughput. While a low value ensures that the expected waiting time to send a CAP message is short. Once DSME is configured according to the needs of an application there are only two different possible values for the fraction \gls{tau} and these cannot be changed at run-time. In this paper, we proposed \ACAP and \DCAP as two extensions of DSME that allow to adopt the fraction \gls{tau} to the current traffic pattern. We verified theoretically and through simulations that both provide a high degree of
responsiveness to traffic fluctuations while keeping the throughput
high. While the first proposal can be implemented within the original
specification the second one requires a deeper intervention. We believe that
with these extensions more demanding IoT applications can be realized
with IEEE 802.15.4.

The proposed mechanisms \ACAP and \DCAP provide the means to build IoT 
applications that can support fluctuating traffic including bursts. To
fully exploit the possibilities they provide, a powerful dynamic
scheduler is needed. While openDSME already provides such a scheduler
which is based on an exponentially weighted moving average filter to
estimate the required GTSs with respect to the traffic demand per
link, it remains to develop {\em Minimal Scheduling Functions} (MSF)
as defined in \cite{Chang:2019} by the 6TiSCH standardization group for TSCH. MSF is designed for best-effort traffic, where most traffic
is periodic monitoring, with occasional bursts. We believe that the extensions proposed in this paper can be leveraged by future Minimal Scheduling Functions for DSME.

\bibliographystyle{splncs04}
\bibliography{document}

\newpage
\appendix
\section{Analysis of Metrics}
Two main metrics are
\begin{enumerate}
	\item Adaptability to time varying traffic expressed as the expected time to send a CAP message on time slot level
	\item Fraction \gls{tau} of CFP's time slots in a dataframe 
\end{enumerate}

Since the periodicity of the DSME frame structure is defined in multiples of a beacon interval (\BI) (e.g. Period of \ACAP, $T_{\ACAP} = 2$\BI), we analyse these metrics in a period interval for each operating mechanism (i.e. \NCAPR, \CAPR, \ACAP and \DCAP). 

\subsection{Definitions}
\begin{figure}[h!]
	\centering
	\includegraphics[width=\textwidth]{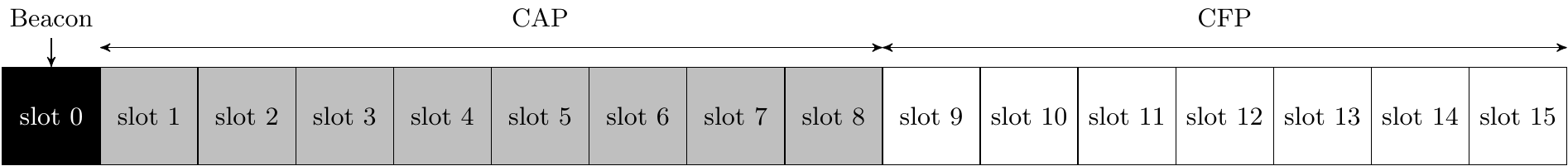}			
	\caption{DSME superframe structure with \NCAPR}
	\label{fig:dsme_superframe_structure}
	
\end{figure}

\begin{figure}[h!]
	\centering
	\includegraphics[width=\textwidth]{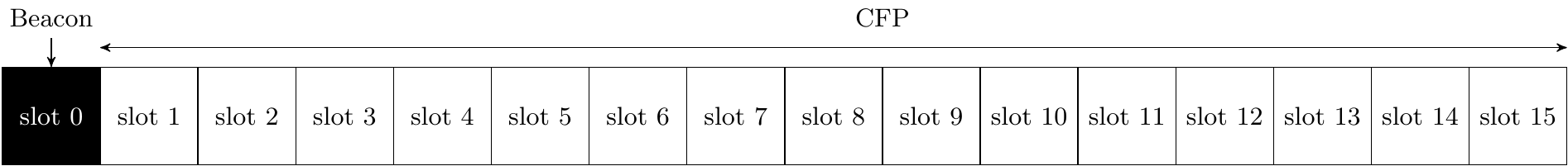}		
	\caption{DSME superframe structure with \CAPR}
	\label{fig:dsme_superframe_structure_with_capr} 
\end{figure}

Let consider the following superframe structures for a network operating in \NCAPR mode (Fig.~\ref{fig:dsme_superframe_structure}) and \CAPR mode (Fig.~\ref{fig:dsme_superframe_structure_with_capr}). The former applies for all superframes of the DSME dataframe structure contained in a \BI. The latter corresponds to superframes different to the first \SF of the first \MSF in a dataframe structure. The excluded \SF follows the structure depicted in Fig.~\ref{fig:dsme_superframe_structure}.  

\begin{itemize}
	\item $ \gls{NTS_SF}=16$: number of time slots per superframe
	\item $\gls{NSF}=2^{\gls{MO}-\gls{SO}}$: number of superframes per multi-superframe
	\item $\gls{NSF_BI}=2^{\gls{BO}-\gls{SO}}$: number of superframes per beacon interval
	\item $\gls{NMSF}=2^{\gls{BO}-\gls{MO}}$: number of multi-superframes per beacon interval
	\item $\gls{SGTS_BI}$: number of GTSs per beacon interval
	\item $\gls{SGTS_SF}$: number of GTSs per superframe

\end{itemize} 

\subsection{Adaptability to time varying traffic on time slot level}\label{sec:appendix:channel_access_time}

This metric can be characterized in terms of the expected number of time slots to send a CAP message, i.e. \gls{TSLOT}. This is, the average waiting time on time slot level to (de)allocate a new GTS. A simple model is as follows: If
a node wishes to allocate a new GTS at the beginning of slot number
$i$ of a \MSF, how many slots must the node wait to reach a CAP slot?. In the following, the formulation for \NCAPR, \CAPR, and \ACAP is presented.

\subsubsection{\NCAPR.}\label{sec:appendix:ch_time_ncapr}

	For each beacon in a \SF, the waiting time is one time slot. Therefore, for all beacons in a \BI the total waiting time is
	\begin{equation} \label{eq:t11}
	t_{1} = \gls{NSF_BI}\times 1
	\end{equation}
	For the CFP time slots in a \SF, the waiting time is 
	$$\sum_{i=9}^{\gls{NTS_SF}-1}{\gls{NTS_SF}+1-i},$$
	and then the total waiting time for all CFP time slots in a 
	\BI is: \begin{eqnarray} \label{eq:t12}
	\nonumber
	t_{2} & = & \gls{NSF_BI}\times \sum_{i=9}^{\gls{NTS_SF}-1}{(\gls{NTS_SF}+1-i)} \\
	& = & \gls{NSF_BI}\times \sum_{i=2}^{\gls{NTS_SF}-8}{i}
	\end{eqnarray}
	From (\ref{eq:t11}) and (\ref{eq:t12}):
	\begin{equation} \label{eq:t1}
	t_{total} = t_{1}+t_{2} =\gls{NSF_BI}\times \sum_{i=1}^{\gls{NTS_SF}-8}{i} = \frac{\gls{NSF_BI}(\gls{NTS_SF}-8)(\gls{NTS_SF}-7)}{2}
	\end{equation}
	
	Thus, the average waiting time over a \BI is:
	\begin{eqnarray} \label{eq:tavg1}
	\nonumber
	t_{\mathrm{avg}} & = & \frac{t_1}{\gls{NSF_BI}\times \gls{NTS_SF}}\\
	\nonumber
	& = & \frac{\gls{NSF_BI}}{\gls{NSF_BI} \times \gls{NTS_SF}}\cdot\frac{(\gls{NTS_SF}-8)(\gls{NTS_SF}-7)}{2}\\
	& = & \frac{(\gls{NTS_SF}-8)(\gls{NTS_SF}-7)}{2\gls{NTS_SF}} 
	\end{eqnarray}
	
	The only influencing parameter is \gls{SO}. The following table exemplifies the waiting time in a superframe structure as shown in Fig.\ref{fig:superframe_structure}.  We can focus on a single \SF, because they are repeated over a \BI \\[1mm]

\setlength\tabcolsep{0.35em}
\begin{tabular}{l|c|c|c|c|c|c|c|c|c|c|c|c|c|c|c|c|c|c|c|} 
Slot-number& 0 & 1 & 2  &3 & 4 & 5 & 6 & 7 & 8 & 9 & 10 & 11 & 12 & 13 & 14 & 15& 16 \\
\hline
Waiting time &1 & 0 & 0& 0 & 0& 0 & 0& 0 & 0&  8& 7 & 6& 5 & 4& 3&2&1 \vspace*{2mm}
\end{tabular}

\noindent
\begin{tabular}{l|c|c|c|c|c|c|c|c|c|c|c|c|c|c|c|} 
Slot-number& 17 & 18  &19 & 20 & 21 & 22 & 23 & 24 & 25 & 26 & 27 & 28 & 29 & 30 & 31 \\
\hline
Waiting time & 0 & 0& 0 & 0& 0 & 0& 0 & 0&  8& 7 & 6& 5 & 4& 3 &2\vspace*{2mm}
\end{tabular}
	
	Hence, the expected waiting time is $72/32= 2.25$ slots as computed in (\ref{eq:tavg1}).\\
\subsubsection{\CAPR.}\label{sec:appendix:ch_time_capr}
	
	Only the first \SF in a \MSF has CAP and the waiting time for its beacon is one time slot. Since there are $\gls{NMSF}$ \textit{MSF}s, the total waiting time is:
	\begin{equation} \label{eq:t21}
	t_{1} = \gls{NMSF}\times 1
	\end{equation}
	
	For the other time slots, in all $\gls{NMSF}$  \textit{MSF}s. the waiting time amounts to:
	\begin{eqnarray} \label{eq:t22}
	\nonumber
	t_{2} & = & \gls{NMSF}\times \sum_{i=9}^{\gls{NSF}\times N-1}{(\gls{NSF}\times \gls{NTS_SF}+1-i)} \\
	& = & \gls{NMSF}\times \sum_{i=2}^{\gls{NSF} \times  \gls{NTS_SF}-8}{i}
	\end{eqnarray}
	From (\ref{eq:t21}) and (\ref{eq:t22}):
	\begin{equation} \label{eq:t2}
	t_{total} = t_{1}+t_{2} =\gls{NMSF}\times \sum_{i=1}^{\gls{NSF} \times  \gls{NTS_SF}-8}{i} = \frac{\gls{NMSF}(\gls{NSF}\times  \gls{NTS_SF}-8)(\gls{NSF}\times  \gls{NTS_SF}-7)}{2}
	\end{equation}
	
	From (\ref{eq:t1}) and (\ref{eq:t2}) the average waiting time over a \BI is:
	\begin{eqnarray} \label{eq:tavg}
	\nonumber
	t_{\mathrm{avg}} & = & \frac{t_1+t_2}{\gls{NSF_BI}\times  \gls{NTS_SF}}\\
	\nonumber
	& = & \frac{\gls{NMSF}}{\gls{NSF_BI} \times  \gls{NTS_SF}}\frac{(\gls{NSF}\times  \gls{NTS_SF}-8)(\gls{NSF}\times  \gls{NTS_SF}-7)}{2} \\
	& = &\frac{(\gls{NSF}\cdot  \gls{NTS_SF}-8)(\gls{NSF}\cdot  \gls{NTS_SF}-7)}{2\gls{NSF}\cdot \gls{NTS_SF}} 
	\end{eqnarray}
	Since $ \gls{NTS_SF}=16$:
	\begin{equation} \label{eq:tavg2} 
	t_{\mathrm{avg}} = 2^{\mathrm{MO-SO+3}}+1.75 / 2^{\mathrm{MO-SO}}-7.5  \mbox{\space(slots)}
	\end{equation}
	

	For the example shown in Fig.\ref{fig:superframe_structure_capOn} with $MO-SO=1$ the individual
	waiting times are as follows\\[1mm]
	
\begin{tabular}{l|c|c|c|c|c|c|c|c|c|c|c|c|c|c|c|c|c|}  
Slot-number& 0 & 1 & 2  &3 & 4 & 5 & 6 & 7 & 8 & 9 & 10 & 11 & 12 & 13 & 14 & 15& 16 \\
\hline
Waiting time &1 & 0 & 0& 0 & 0& 0 & 0& 0 & 0& 24&23&22& 21& 20 & 19& 18 & 17  \vspace*{2mm}
\end{tabular}

\noindent
\begin{tabular}{l|c|c|c|c|c|c|c|c|c|c|c|c|c|c|c|c|c|c|}  
Slot-number& 17 & 18  &19 & 20 & 21 & 22 & 23 & 24 & 25 & 26 & 27 & 28 & 29 & 30 & 31 \\
\hline
Waiting time & 16&15 & 14 & 13& 12 & 11& 10 & 9& 8 & 7& 6 & 5& 4 & 3& 2  \vspace*{2mm}
\end{tabular}   
	
	Then, the expected waiting time is $300/32= 9.375$ slots, that corresponds with the expression found in (\ref{eq:tavg2}).\\
\subsubsection{\ACAP.}\label{sec:appendix:ch_time_acap}
	
	In this case a period equals $2$\BI. Then, the expected time corresponds to an average of the waiting time for \CAPR (\ref{eq:t1}) and \NCAPR (\ref{eq:t2}) in a \BI. Then, the total waiting time is:
	
	\begin{equation} \label{eq:t3}
	t_{total} = t_{NCR}+t_{CR} =\gls{NMSF}\times \sum_{i=1}^{\gls{NSF} \times  \gls{NTS_SF}-8}{i} = \frac{\gls{NMSF}(\gls{NSF}\times  \gls{NTS_SF}-8)(\gls{NSF}\times  \gls{NTS_SF}-7)}{2}
	\end{equation}

	From (\ref{eq:t3}) the average waiting time is
	
	\begin{eqnarray} \label{eq:tavg3}
	\nonumber
	t_{\mathrm{avg}} & = & \frac{t_1+t_2}{2\times \gls{NSF_BI}\times \gls{NTS_SF}}\\
	\nonumber
	& = & \frac{\gls{NSF_BI}}{2\times \gls{NSF_BI} \times  \gls{NTS_SF}}\frac{( \gls{NTS_SF}-8)( \gls{NTS_SF}-7)}{2} + \frac{\gls{NMSF}}{2\times \gls{NSF_BI} \times  \gls{NTS_SF}}\frac{(\gls{NSF}\times  \gls{NTS_SF}-8)(\gls{NSF}\times  \gls{NTS_SF}-7)}{2} \\
	& = & \frac{( \gls{NTS_SF}-8)( \gls{NTS_SF}-7)}{4 \gls{NTS_SF}} + \frac{(\gls{NSF}\cdot  \gls{NTS_SF}-8)(\gls{NSF}\cdot  \gls{NTS_SF}-7)}{4\gls{NSF}\cdot  \gls{NTS_SF}} 
	\end{eqnarray}
	Since $ \gls{NTS_SF}=16$:
	\begin{equation} \label{eq:tavg3}
	t_{\mathrm{avg}} = 2^{\mathrm{\gls{MO}-\gls{SO}+2}}+0.875/ 2^{\mathrm{\gls{MO}-\gls{SO}}}-2.625 \mbox{\space(slots)}
	\end{equation}
	

	For the example in Fig.~\ref{fig:superframe_structure_HCAP} of \ACAP the expected waiting time is $5.8125$ slots, this value matches with the formula calculated in (\ref{eq:tavg3}).
	
	
	For the variant \DCAP the computation is a bit more difficult, given the dynamism that this mechanism uses. This is, the number of GTSs is changing over time.
	
	The unit for the expected waiting time is time slots, depending on the value of \gls{SO} this can be converted into ms. This calculation does not respect the fact, that a slot negotiation takes more than one time slot, i.e., if a slot allocation is triggered at the beginning of the last CAP's time slot, then it will take more time compared to triggering at the beginning of the first CAP's time slot. Respecting this fact may lead to very difficult computations.

\subsection{Fraction \gls{tau} of CFP's time slots in a dataframe }

This metric is defined as the fraction of CFP's time slots in a dataframe, which is bounded in a \BI.  In the following, the formulation for \NCAPR, \CAPR, and \ACAP is presented.

\subsubsection{\NCAPR.}
	
	For each superframe the total number of GTSs is
	
	\begin{eqnarray} \label{eq:gts_s}
	\nonumber
	\gls{SGTS_SF} & = &\sum_{i=9}^{ \gls{NTS_SF}-1}{\mbox{slot}_i}\\
	\nonumber
	& = &  \gls{NTS_SF}-9
	\end{eqnarray}
	Given that the frame structure is the same in all superframes. The total number of GTS per \BI is calculated as
	
	\begin{eqnarray} \label{eq:gts_bi}
	\nonumber
	\gls{SGTS_BI} & = & \gls{NSF_BI}\times \gls{SGTS_SF}\\
	\nonumber
	& = & \gls{NSF_BI}\times ( \gls{NTS_SF}-9)
	\end{eqnarray}
	
	Thus, the fraction of CFP's time slots in a \BI is estimated as
	
	\begin{eqnarray} \label{eq:net_dsme}
	\nonumber
	\gls{tau} & = & \frac{\gls{SGTS_BI}}{\gls{NSF_BI}\times  \gls{NTS_SF}}\\
	\nonumber
	& = & \frac{\gls{NSF_BI}\times ( \gls{NTS_SF}-9)}{\gls{NSF_BI}\times \gls{NTS_SF}}\\
	& = & \frac{( \gls{NTS_SF}-9)}{ \gls{NTS_SF}}
	\end{eqnarray}
	
	Since $\gls{NTS_SF}=16$, $\gls{tau} = 7/16=0.4375$ and it is independent of values of $SO,MO,BO$.
	
\subsubsection{\CAPR.}
	
	For CFP's time slots, the baseline is the number of GTSs per \MSF. Then, the total number of GTSs per \BI amounts to:
	
	\begin{eqnarray} \label{eq:gts_bi_capr}
	\nonumber
	\gls{SGTS_BI} & = & \gls{NMSF}\times (\sum_{i=9}^{\gls{NSF}\times  \gls{NTS_SF}-1}{\mbox{slot}_i} - (\gls{NSF}-1)\times \mbox{slot}_{0}) \\
	\nonumber
	& = & \gls{NMSF}\times (\gls{NSF}\times( \gls{NTS_SF}-1)-8)
	\end{eqnarray}
	
	The calculation of the fraction \gls{tau} is
	
	\begin{eqnarray} \label{eq:net_capr}
	\nonumber
	\gls{tau} & = & \frac{\gls{SGTS_BI}}{\gls{NSF_BI}\times  \gls{NTS_SF}}\\
	\nonumber
	& = & \frac{\gls{NMSF}\times (\gls{NSF}\times( \gls{NTS_SF}-1)-8)}{\gls{NSF_BI}\times  \gls{NTS_SF}}\\
	& = & \frac{( \gls{NTS_SF}-1)-8\times2^{\mathrm{\gls{SO}-\gls{MO}}}}{ \gls{NTS_SF}}
	\end{eqnarray}

	Since $ \gls{NTS_SF}=16$:
	\begin{equation} \label{eq:net_capr_2}
	\gls{tau} = 0.9375- 2^{\mathrm{(\gls{SO}-\gls{MO}-1)}}
	\end{equation}
	
\subsubsection{\ACAP.}
	
	This estimation is made by calculating the average of the fraction \gls{tau} for \NCAPR and  \CAPR in $2$\BI. Then, the fraction of CFP's time slots in a dataframe is
	
	\begin{eqnarray} \label{eq:net_capr}
	\nonumber
	\gls{tau} & = & \frac{\gls{NSF_BI}\times ( \gls{NTS_SF}-9)+ \gls{NMSF}\times (\gls{NSF}\times( \gls{NTS_SF}-1)-8)}{2\times \gls{NSF_BI}\times  \gls{NTS_SF}}\\
	\nonumber
	& = & \frac{\gls{NSF_BI}\cdot((\gls{NTS_SF}-9)+ 2^{\mathrm{(\gls{SO}-\gls{MO})}}\times (\gls{NSF}\times( \gls{NTS_SF}-1)-8)}{2\times \gls{NSF_BI}\times  \gls{NTS_SF}}\\  
	\nonumber
	& = & \frac{2 \gls{NTS_SF}-10-2^{\mathrm{(\gls{SO}-\gls{MO}}+3)}}{2 \gls{NTS_SF}}  
	\end{eqnarray}

Since  $ \gls{NTS_SF}=16$:
\begin{equation} \label{eq:net_capr_2}
\gls{tau} = 0.6875 - 2^{\mathrm{(\gls{SO}-\gls{MO}-2)}}
\end{equation}

\end{document}